\documentstyle[preprint,aps]{revtex}
\tighten
\draft
\widetext
\input epsf
\preprint{HUTP-99/A022, NUB 3198, EFI-99-15}
\bigskip
\bigskip
\begin{document}
\title{Type I on (Generalized) Voisin-Borcea Orbifolds and 
Non-perturbative Orientifolds}
\medskip

\author{Zurab Kakushadze\footnote{E-mail: 
zurab@string.harvard.edu}}

\bigskip
\address{Jefferson Laboratory of Physics, Harvard University,
Cambridge,  MA 02138\\
and\\
Department of Physics, Northeastern University, Boston, MA 02115\\
and\\
Enrico Fermi Institute, University of Chicago, Chicago, IL 60637}

\date{April 29, 1999}
\bigskip
\medskip
\maketitle

\begin{abstract}
{}We consider non-perturbative four dimensional ${\cal N}=1$ space-time
supersymmetric orientifolds corresponding to Type I compactifications
on (generalized) Voisin-Borcea orbifolds.
Some states in such compactifications arise in
``twisted'' open string sectors which lack world-sheet description in terms
of D-branes. Using Type I-heterotic duality as well as the map between 
Type IIB orientifolds and F-theory we are able to obtain the
massless spectra of such orientifolds. The four dimensional
compactifications we discuss in this context are examples of chiral 
${\cal N}=1$ supersymmetric 
string vacua which are non-perturbative from both orientifold
and heterotic points of view. In particular, they contain 
both D9- and D5-branes
as well as non-perturbative ``twisted'' open string sector states.
We also explain the origins of various inconsistencies arising in such 
compactifications for certain choices of the gauge bundle.

\end{abstract}

\section{Introduction}

{}In the recent years much progress has been made in understanding six and
four dimensional
orientifold compactifications. Various six dimensional orientifold vacua
were constructed, for instance, in \cite{PS,GP,GJ,bij}. Generalizations of
these constructions to four dimensional orientifold vacua have also been
discussed in detail in \cite{BL,Sagnotti,ZK,KS,Zw,ibanez,KST,blumen,223}. 
In many cases the perturbative world-sheet approach to orientifolds gives 
rise to consistent anomaly free vacua in six and four dimensions. 
However, as was
pointed out in \cite{KST}, there
are cases where the perturbative orientifold description is inadequate as it
misses certain
non-perturbative sectors giving rise to massless states.
In certain cases this inadequacy results in obvious
inconsistencies such as lack of tadpole and anomaly cancellation. Examples
of such cases were discussed in \cite{Zw,ibanez,KST}. In other cases,
however, the issue is more subtle as the non-perturbative states arise in
anomaly free combinations, so that they are easier to miss. 

{}Thus, let us consider Abelian $T^6/\Gamma$ orbifold compactifications of
Type I with ${\cal N}=1$ supersymmetry in four dimensions. 
The requirement that the orbifold group $\Gamma$ act 
crystallographically on $T^6$ restricts the allowed choices of $\Gamma$ to
${\bf Z}_3^*$, ${\bf Z}_4^*$, ${\bf Z}_6^*$, ${\bf Z}_6^\prime$, ${\bf Z}_7$,
${\bf Z}_8$, ${\bf Z}_8^\prime$, ${\bf Z}_{12}$ and ${\bf Z}_{12}^\prime$ for 
$\Gamma\approx{\bf Z}_N$, and to ${\bf Z}_2\otimes {\bf Z}_2$, 
${\bf Z}_2\otimes {\bf Z}_4$, ${\bf Z}_2\otimes {\bf Z}_6$, ${\bf Z}_2\otimes
{\bf Z}_6^*$, ${\bf Z}_3\otimes
{\bf Z}_3$, ${\bf Z}_3\otimes {\bf Z}_6$ and ${\bf Z}_4\otimes {\bf Z}_4$
for $\Gamma\approx{\bf Z}_N\otimes {\bf Z}_M$. Here we use star and prime
to distinguish different cyclic groups of the same order which act differently
on $T^6$ (the precise actions of these orbifold groups on $T^6$ are given in 
the subsequent sections). In some of the above cases  
Type I-heterotic duality enables one to argue along the lines of 
\cite{ZK,KS,KST} that the non-perturbative states become heavy 
and decouple once the 
corresponding orbifold singularities are appropriately blown
up. In particular, these are the ${\bf Z}_3^*$ \cite{Sagnotti,ZK}, 
${\bf Z}_7$, ${\bf Z}_3\otimes {\bf Z}_3$ and ${\bf Z}_6^*$ \cite{KS},
and ${\bf Z}_2\otimes {\bf Z}_6^*$ \cite{223}
orbifold cases. As was argued in \cite{KST}, in other perturbatively tadpole
free orientifolds, such as the ${\bf Z}_6^\prime$, 
${\bf Z}_2\otimes{\bf Z}_6$,     
${\bf Z}_3\otimes{\bf Z}_6$ and ${\bf Z}_6\otimes{\bf Z}_6$ \cite{Zw}, and 
${\bf Z}_{12}$ \cite{ibanez} 
orbifold cases, the non-perturbative open string sectors do give rise to 
massless states which do not decouple even for blown-up orbifolds.
Recently some examples of non-perturbative orientifolds were explicitly
constructed in \cite{NP}. Thus, the four dimensional example discussed
in \cite{NP} is based on the ${\bf Z}_6^\prime$ orbifold.
The purpose of this paper, which is the follow-up
of \cite{NP}, is to extend the discussions of \cite{NP} to understand
all other orientifold cases as well. 
In particular, we will discuss the ${\bf Z}_2\otimes{\bf Z}_6$,     
${\bf Z}_3\otimes{\bf Z}_6$, ${\bf Z}_6\otimes{\bf Z}_6$ and 
${\bf Z}_{12}$ cases. The first three cases lead to anomaly free 
non-perturbative orientifolds. The last case, however, turns out to suffer from
a non-perturbative anomaly whose origins we explain in detail in section V.  
Moreover, we elaborate on the rest of the orbifold compactifications mentioned
above, namely, the ${\bf Z}_2\otimes {\bf Z}_4$ and ${\bf Z}_4\otimes
{\bf Z}_4$ \cite{Zw}, and ${\bf Z}_8$, ${\bf Z}_8^\prime$ and
${\bf Z}_{12}$ \cite{ibanez} cases, which 
where shown to suffer from perturbative tadpoles in \cite{Zw} and 
\cite{ibanez}, respectively. In \cite{KST} this was explained by considering
F-theory duals of these compactifications. In this paper we elaborate the 
arguments of \cite{KST}. In particular, we point out that the corresponding
inconsistencies arise due to the particular choices of the gauge bundle
in the models of \cite{Zw,ibanez} (these gauge bundles are perturbative
from the orientifold viewpoint).

{}The origin of non-perturbative states in orientifold compactifications
can already be understood in six dimensions. Thus, in the K3 orbifold examples 
of \cite{GJ} the orientifold projection is not $\Omega$, which we will use
to denote that in the smooth K3 case, but rather $\Omega J^\prime$, where
$J^\prime$ maps the $g$ twisted sector to its conjugate $g^{-1}$ twisted
sector (assuming $g^2\not=1$) \cite{pol}. Geometrically this can be viewed
as a permutation of two ${\bf P}^1$'s associated with each fixed point of
the orbifold \cite{KST}. (More precisely, these ${\bf P}^1$'s correspond to
the orbifold blow-ups.) This is different from the orientifold projection
in the smooth case where (after blowing up) the orientifold projection
does {\em not} permute the two ${\bf P}^1$'s. In the case of the 
$\Omega J^\prime$ projection the ``twisted'' open string sectors corresponding
to the orientifold elements $\Omega J^\prime g$ are absent
\cite{blum,KST}. However, if the orientifold projection is $\Omega$, then
the ``twisted'' open string sectors corresponding
to the orientifold elements $\Omega g$ are present \cite{KST}. 
In fact, these states
are non-perturbative from the orientifold viewpoint and are required for
gravitational anomaly cancellation in six dimensions. In
certain cases Type I-heterotic duality allows one to understand such
sectors and construct the corresponding models explicitly \cite{NP}.

{}In four dimensional orientifolds with ${\cal N}=1$ supersymmetry there
are always sectors (except in the ${\bf Z}_2\otimes {\bf Z}_2$ model of
\cite{BL} which is completely perturbative from the orientifold viewpoint)
such that there is only one ${\bf P}^1$ per fixed point, so only the
$\Omega$ orientifold projection is allowed. This results in
non-perturbative ``twisted'' open string sectors, which, as we have
already mentioned, decouple in certain cases once the appropriate blow-ups 
are performed. In other cases we must include these states to obtain the
complete description of a given orientifold.      

{}Some of the non-perturbative orientifolds have perturbative heterotic
duals. However, non-perturbative orientifolds with, say, D5-branes are
non-perturbative from the heterotic viewpoint. In this paper we are
therefore exploring some vacua in the region ${\cal D}$ in Fig.1 (which has
been borrowed from the second reference in \cite{223}). 
Thus, the ${\bf Z}_6^\prime$ orbifold
compactification discussed in \cite{NP}, as well as the ${\bf Z}_2\otimes
{\bf Z}_6$ orbifold compactification we discuss in section V of this paper,
are examples of four
dimensional {\em chiral} ${\cal N}=1$ supersymmetric string vacua which are
non-perturbative from both orientifold and heterotic points of view.

{}The key point which allows us to understand four dimensional 
non-perturbative orientifolds discussed in this paper 
is the fact that these models correspond to Type I
compactifications on generalized Voisin-Borcea orbifolds of the form
$(T^2\otimes {\mbox{K3}})/{\bf Z}_N$ ($N=2,3,4,6$). One of the  simplifying 
features of
these compactifications is that once non-perturbative K3 orientifolds are 
understood (and this is partially facilitated by the fact that in six 
dimensions the gravitational anomaly cancellation condition is rather
constraining), it is possible to extend the corresponding results to 
generalized Voisin-Borcea compactifications. Another nice feature is that,
by T-dualizing along $T^2$, we can map these modes to 
F-theory compactifications on the corresponding Calabi-Yau four-folds, and
the (at least partially) geometric picture arising in the F-theory context
is very helpful in understanding such vacua.    

{}The rest of this paper is organized as follows. In section II
we use the map between Type I compactifications on K3 and F-theory on
Voisin-Borcea orbifolds to determine which gauge bundles are perturbative from
the orientifold viewpoint. In section III we give explicit models corresponding
to K3 compactifications of Type I with these perturbative gauge bundles. 
In section IV we move down to four dimensions and consider Type I 
compactifications on Voisin-Borcea orbifolds. In particular, using the map
between such Type I vacua and F-theory on Calabi-Yau four-folds we give the
constraints necessary for tadpole cancellation in the cases where
the gauge bundle is perturbative. In particular, this discussion relies on 
the results of section II. In section V we discuss Type I 
compactifications on generalized Voisin-Borcea orbifolds 
$(T^2\otimes {\mbox{K3}})/{\bf Z}_N$ with $N=3,4,6$. In particular, we argue
that the ${\bf Z}_{12}$ case suffers from a non-perturbative anomaly.
In section VI we discuss some directions for extending our results. In 
particular, we consider Type I on K3 with non-zero NS-NS $B$-field and the
gauge bundles with vector structure.

\section{Type I on K3 (Orbifolds)}

{}In this section we discuss six dimensional ${\cal N}=1$ supersymmetric
Type I compactifications on K3. In particular,
what we would like to understand here is which choices of the gauge bundle
are perturbative from the orientifold viewpoint. Our discussion here applies
to both Type I and ${\mbox{Spin}}(32)/{\bf Z}_2$ heterotic compactifications 
on K3, which is due to Type I-heterotic duality. The dictionary between the 
two descriptions includes mapping 5-branes on the Type I side, which are
made of some number of D5-branes, to NS 5-branes (or, equivalently, small 
instantons) on the heterotic side. 

{}Thus, consider Type I on a K3 surface. In order to determine a particular 
ground state
of the theory, we must specify not only the compactification space, 
which in this case is K3, but also the choice of the gauge bundle. Thus, 
in the orientifold description of Type I non-perturbative effects can arise 
from two different sources: (1) the perturbative orientifold description may 
fail to capture all states arising due to, say, various singularities in K3
itself; (2) a given choice of the gauge bundle may not have perturbative 
orientifold description. The second source of non-perturbative physics is 
generally more non-trivial to understand. Here, however, we will be interested
in determining which choices of the gauge bundle do possess perturbative
orientifold description, and once we understand this class of gauge bundles, we
can then focus on the first source of non-perturbative effects which are easier
to handle due to various hints from geometry.

{}The case of K3 is particularly convenient as (unlike in the case of 
Calabi-Yau three-folds) topologically we are dealing with one surface. This
enables us to understand perturbative gauge bundles in the K3 case by going to
orbifold limits which have a simple description. Thus, let us consider orbifold
K3's of the form ${\mbox{K3}}=T^4/{\bf Z}_N$, where $N=2,3,4,6$ are the choices
allowed by the requirement that the orbifold act crystallographically on the
four-torus $T^4$. Let $z_1,z_2$ be the complex coordinates parametrizing 
$T^4$. Then the action of the generator $g$ of ${\bf Z}_N$ is given by
\begin{equation}
 g z_1=\omega z_1~,~~~gz_2=\omega^{-1} z_2~,
\end{equation}   
where $\omega\equiv\exp(2\pi i/N)$.

{}Next, consider Type I on $T^4/{\bf Z}_N$. We will assume that the (untwisted)
NS-NS $B$-field is trivial. Then we can view Type I on $T^4/{\bf Z}_N$ as the
$\Omega$ orientifold of Type IIB on $T^4/{\bf Z}_N$. (More precisely, the 
$\Omega$ orientifold is well defined after the appropriate blow-ups of the 
orbifold singularities other than ${\bf Z}_2$.) The theory has ${\cal N}=1$
supersymmetry in six dimensions. The massless spectrum of the closed string 
sector consists of the gravity supermultiplet, one tensor supermultiplet, and
20 hypermultiplets (which are neutral under the non-Abelian gauge symmetries
in the open string sector). In the open string sector we have 32 D9-branes, 
which is required by the untwisted tadpole cancellation conditions. In the 
${\bf Z}_3$ case we do not have any D5-branes. In the ${\bf Z}_2,{\bf Z}_4,
{\bf Z}_6$ cases tadpole cancellation conditions also 
require introduction of 32 D5-branes. 

{}To understand the gauge bundles in all of the above cases, recall that in
the case of K3 anomaly cancellation requires that the number $n_I$ of 
instantons embedded in the gauge bundle and the number $n_5$ of 5-branes 
(that is, small instantons) must add up to 24: $n_I+n_5=24$. 
In the ${\bf Z}_3$ all 24 instantons are embedded in the gauge bundle, so the
number of 5-branes is zero. This gauge bundle is therefore perturbative from 
the heterotic viewpoint. On the orientifold side it is defined by the action
of the orbifold group on the Chan-Paton charges which is given by the 
following $16\times 16$ matrix\footnote{Throughout this paper 
(unless specified otherwise) we work with
$16\times 16$ (rather than $32\times 32$) Chan-Paton matrices for we choose
not to count the orientifold images of the corresponding D-branes.} \cite{GJ}:
\begin{equation}\label{Z_3}
 {\bf Z}_3:~~~\gamma_{g}=
 {\mbox{diag}}(\alpha {\bf I}_4,\alpha^{-1} {\bf I}_4,{\bf I}_8)~,
\end{equation}
where $\alpha\equiv\exp(2\pi i/3)$, and ${\bf I}_n$ denotes the $n\times n$ 
identity matrix.

{}Next, consider the ${\bf Z}_2$ case. Here we have 32 D5-branes which form 
$n_5=8$
dynamical 5-branes. Here two pairings take place - one due to the orientifold
projection, and the other one due the ${\bf Z}_2$ orbifold projection. We 
therefore conclude that $n_I=16$ in this case. This gauge bundle is 
non-perturbative from the heterotic viewpoint. On the orientifold side the
corresponding Chan-Paton matrix is given by \cite{PS,GP}:
\begin{equation}\label{Z_2}
 {\bf Z}_2:~~~\gamma_{g}={\mbox{diag}}(i{\bf I}_8,-i{\bf I}_8)~.
\end{equation}

{}As to the ${\bf Z}_4$ and ${\bf Z}_6$ cases, the corresponding models
are on the same moduli as the ${\bf Z}_2$ model. Indeed, in all three of 
these cases the gauge group can be Higgsed completely. In fact, in the 
${\bf Z}_4$ and ${\bf Z}_6$ cases this can already be done just with the
matter arising in the sectors which are perturbative from the orientifold 
viewpoint \cite{GJ1}. 
Then the gravitational anomaly cancellation condition implies that
after Higgsing the number of hypermultiplets in the open string sector is the 
same in all three models (the closed string spectra of these models are 
identical), hence they are on the same moduli. This implies that in the ${\bf
Z}_4$ and ${\bf Z}_6$ cases we also have $n_5=8$ and $n_I=16$. In the 
orientifold language the corresponding Chan-Paton matrices are given by 
\cite{GJ}:
\begin{eqnarray}\label{Z_4}
 &&{\bf Z}_4:~~~\gamma_{g}={\mbox{diag}}(\beta {\bf I}_4,-\beta {\bf I}_4,
 \beta^{-1}{\bf I}_4,-\beta^{-1}{\bf I}_4)~,\\
 \label{Z_6}
 &&{\bf Z}_6:~~~\gamma_{g}={\mbox{diag}}(i\alpha {\bf I}_2,-i\alpha {\bf I}_2,
 i\alpha^{-1}{\bf I}_2,-i\alpha^{-1}{\bf I}_2, i{\bf I}_4,-i{\bf I}_4)~,
\end{eqnarray} 
where $\beta\equiv\exp(\pi i/4)$.

{}Thus, the gauge bundles that are perturbative from the orientifold viewpoint
have $(n_5,n_I)=(0,24)$ and $(n_5,n_I)=(8,16)$. It is instructive to understand
these gauge bundles a bit better. In particular, here we would like to discuss
them in the heterotic and F-theory pictures. On the heterotic side there are 
three topologically distinct gauge bundles with 
$(n_5,n_I)=(0,24)$. These are characterized by the generalized second 
Stiefel-Whitney class ${\widetilde w}_2$ (which is an element of 
$H^2({\mbox{K3}},{\bf Z})$, or, more precisely, of
$H^2({\mbox{K3}},{\bf Z}_2)$ as ${\widetilde w}_2$ is defined modulo a shift
by twice a lattice vector of $H^2({\mbox{K3}},{\bf Z})$) \cite{berkooz,paul}.  
The three distinct gauge bundles correspond to: (1) ${\widetilde w}_2=0$; (2)
${\widetilde w}_2\cdot{\widetilde w}_2=0~({\mbox{mod}}~4)$; (3) 
${\widetilde w}_2\cdot{\widetilde w}_2=2~({\mbox{mod}}~4)$. In the first case
we have vector structure, while (2) and (3) are the cases without vector 
structure.

{}To understand the above three cases better, it is useful to think about
the corresponding gauge bundles in terms of the ${\bf Z}_2$ orbifold action
on the ${\mbox{Spin}}(32)/{\bf Z}_2$ degrees of freedom in the heterotic 
context. This action can be viewed in terms of a ${\bf Z}_2$ valued shift $V$ 
of the ${\mbox{Spin}}(32)/{\bf Z}_2$ lattice. 
Let us use the Cartan basis of $SO(32)$ to 
write $V$. If $2V$ belongs to the $SO(32)$ lattice, 
then we have vector 
structure, and ${\widetilde w}_2=0$. Two inequivalent gauge bundles of this 
type which are perturbative from the heterotic viewpoint are given by
$V=((1/2)^2~0^{14})$ and $V=((1/2)^6~0^{10})$ (here we have chosen the 
surviving spinor of ${\mbox{Spin}}(32)/{\bf Z}_2$ to be given by 
$((1/2)^{16})$ modulo $SO(32)$ shifts). In particular, they satisfy the modular
invariance constraints such as the level matching condition. (Thus,
the gauge bundle given by $V=((1/2)^2~0^{14})$ corresponds to the standard 
embedding for the ${\bf Z}_2$ orbifold.) These gauge bundles are 
non-perturbative from the orientifold viewpoint.

{}However, we can have $V$ such that
$2V$ does not belong to the $SO(32)$ lattice, albeit it does belong to the
${\mbox{Spin}}(32)/{\bf Z}_2$ lattice. Two inequivalent gauge bundles of this
type are given by $V=((1/4)^{16})$ and $V=((1/4)^{15}(3/4))$. Note that
for such gauge bundles ${\widetilde w}_2\not=0$, and ${\widetilde w}_2\cdot
{\widetilde w}_2=(2V)^2$, where by $(2V)^2$ we mean the length squared of the
16-vector $2V$. The gauge bundle given by $V=((1/4)^{16})$ is 
non-perturbative from the heterotic viewpoint - it does not satisfy the level 
matching condition. In fact, this is precisely the gauge bundle in the 
Type IIB orientifold of $T^4/{\bf Z}_2$ \cite{PS,GP}. As we have already 
mentioned, this gauge bundle corresponds to embedding 16 instantons accompanied
by 8 5-branes to cancel the anomalies. On the other hand, the gauge bundle
$V=((1/4)^{15}(3/4))$ is perturbative from the heterotic viewpoint - it 
satisfies the level matching requirement, and corresponds to embedding all 24
instantons. It is, however, non-perturbative from the orientifold viewpoint.

{}Finally, let us discuss the F-theory duals of the above cases. In \cite{SS}
it was shown that in the ${\widetilde w}_2=0$ case the dual F-theory 
compactification is on the Calabi-Yau three-fold given by the $T^2$ fibration
over the base ${\bf F}_4$. (Here ${\bf F}_n$ are Hirzebruch surfaces.) In the
case of the ${\bf Z}_3$ orbifold this can be seen in two ways. First, we can 
Higgs the $U(8)\otimes SO(16)$ gauge group at the orbifold point (see Table I)
down to $SO(8)\otimes SO(8)$ with 10 adjoint hypermultiplets in the first 
$SO(8)$ and no matter in the second $SO(8)$. This is precisely the spectrum 
of the F-theory dual on the Calabi-Yau three-fold with the base ${\bf F}_4$
(at the Voisin-Borcea orbifold point - see section IV). 
On the other hand, we can
explicitly construct the $SO(8)\otimes SO(8)$ point in the string language.
Let K3 be $(T^2\otimes T^2)/{\bf Z}_3$, where ${\bf Z}_3$ simultaneously 
rotates the two $T^2$'s by $2\pi/3$. Then consider the following choice of the
gauge bundle. First, let us turn on a ${\bf Z}_2$ valued 
Wilson line on the $a$-cycle of, say, the first $T^2$ such that it breaks
$SO(32)$ down to $SO(16)\otimes SO(16)$. Next, let us turn on the second
${\bf Z}_2$ valued 
Wilson line on the $b$-cycle of the same $T^2$ such that it further breaks
$SO(16)\otimes SO(16)$ down to $SO(8)^4$. It is not difficult to see that
this theory has ${\bf Z}_3$ symmetry under which, say, the first three 
$SO(8)$'s are cyclically permuted. In fact, since the $T^2$'s are hexagonal,
under the ${\bf Z}_3$ rotation the $a$-cycles maps to the $b$-cycle, and so
the first Wilson line maps to the second Wilson line. The choice of the gauge 
bundle corresponding to the ${\bf Z}_3$ orbifold then precisely consists of    
the twist permuting the first three $SO(8)$'s, while leaving the last $SO(8)$ 
untouched. The resulting model has $SO(8)_3\otimes SO(8)_1$ gauge group, where
the subscript indicates the current algebra level via which the corresponding
subgroup is realized.
The untwisted sector contains the gravity supermultiplet, one tensor 
supermultiplet, 2 neutral hypermultiplets, and one hypermultiplet in the 
adjoint of $SO(8)_3$. The twisted sector contains 18 neutral hypermultiplets
(corresponding to the blow-up modes of the orbifold), and 9 hypermultiplets
in the adjoint of $SO(8)_3$. This spectrum precisely matches that of the 
F-theory dual corresponding to the base ${\bf F}_4$.  

{}As to the cases with 
${\widetilde w}_2\cdot{\widetilde w}_2=0~({\mbox{mod}}~4)$ and  
${\widetilde w}_2\cdot{\widetilde w}_2=2~({\mbox{mod}}~4)$, the corresponding
dual F-theory compactifications are on Calabi-Yau three-folds given by $T^2$
fibrations over ${\bf F}_0$ respectively ${\bf F}_1$ \cite{SS}. Here we note 
that these cases also correspond to Voisin-Borcea orbifolds, but at the latter
points in the respective moduli spaces the corresponding bases are singular
leading to additional matter \cite{MV} precisely such that the resulting
massless spectra match those of the heterotic/Type I duals. Also, as was 
pointed out in \cite{KST}, in the 
${\widetilde w}_2\cdot{\widetilde w}_2=0~({\mbox{mod}}~4)$ case one can also
view the dual F-theory compactification as on a singular Calabi-Yau three-fold 
(with Hodge numbers $(h^{1,1},h^{2,1})=(3,51)$) given by the ``non-geometric''
${\bf Z}_2\otimes {\bf Z}_2$ orbifold of $T^2\otimes T^2\otimes T^2$ with
discrete torsion \cite{VW}. This F-theory compactification is in turn on the 
same moduli as that on a smooth Calabi-Yau three-fold (with Hodge 
numbers $(h^{1,1},h^{2,1})=(3,243)$) given by the $T^2$ fibration over the
base ${\bf P}^1\otimes {\bf P}^1$ (that is, the smooth ${\bf F}_0$ surface)
\cite{KST}.  

{}To summarize, the gauge bundles which are perturbative from the orientifold
viewpoint are those with $(n_5,n_I)=(0,24)$, ${\widetilde w}_2=0$ (this is the
case for the gauge bundle in the ${\bf Z}_3$ case (\ref{Z_3})), and
$(n_5,n_I)=(8,16)$, ${\widetilde w}_2\cdot{\widetilde w}_2=0~({\mbox{mod}}~4)$
(this is the case for the gauge bundles in the ${\bf Z}_2$ (\ref{Z_2}),
${\bf Z}_4$ (\ref{Z_4}), and ${\bf Z}_6$ (\ref{Z_6}) cases).

\section{Non-perturbative K3 Orientifolds}

{}In the previous section we have discussed the gauge bundles in K3 
compactifications of Type I which are perturbative from the orientifold 
viewpoint. In this section we would like to discuss the explicit Type I
vacua arising upon compactifications on K3 with these gauge bundles. These
compactifications can be viewed as the $\Omega$ orientifolds of Type IIB
on (the corresponding orbifold limits of) K3. Thus, the ${\bf Z}_2$ case
with the gauge bundle given by (\ref{Z_2}) (the action of the orbifold group
on both D9- and D5-branes is the same) was discussed in \cite{PS,GP}. This
model is completely perturbative from the orientifold viewpoint. Thus, the
``$\Omega$-twisted'' sector of the orientifold can be viewed as the usual
(``untwisted'') 99 and 55 open string sectors. The ``$\Omega g$-twisted''
sector (with $g^2=1$ in this case) can be viewed as the usual (``untwisted'')
59 open string sector. These sectors are perturbative from the orientifold 
point of view. The massless open spectrum of this model is given in Table I.  

{}In the ${\bf Z}_3$ case we do have non-perturbative sectors, however. Thus,
the ``$\Omega$-twisted'' sector of the orientifold corresponds to the 
untwisted 99 open string sector (there are no D5-branes in this case). The
``$\Omega g$- and $\Omega g^2$-twisted'' sectors (with $g^3=1$ in this case) 
correspond to the ${\bf Z}_3$ 
``twisted'' 99 open string sector. This model has a 
perturbative heterotic dual which enables one to obtain its massless spectrum
\cite{NP}, and the latter is given\footnote{For the sake of simplicity, 
throughout this paper we omit the $U(1)$ charges. In the untwisted open string
sectors they are straightforward to determine from the standard Chan-Paton
charge assignments. In the twisted open string sectors, however, these
$U(1)$ charges are generically fractional (in the normalization where
the fundamental of $SU(N)\subset U(N)$ in the untwisted open string sectors has
the $U(1)$ charge $+1$). In certain cases one can determine these 
charges using Type I-heterotic duality (as on the heterotic side these charges
can be computed perturbatively in the corresponding twisted sectors). In other
cases determining the twisted sector $U(1)$ charges can be more involved. Also,
some of the twisted closed string states can transform non-trivially under 
gauge transformations of some of the $U(1)$'s. The corresponding $U(1)$ charges
(in the appropriate basis) can therefore be non-zero (and sometimes even
fractional).} in Table I. 

{}In the ${\bf Z}_6$ case we do not have a perturbative heterotic dual as the
Type I model contains D5-branes. Nonetheless, it is still possible to 
determine the spectrum of the model. Thus, the ${\bf Z}_3$ twisted 99
open string sector is the same as in the ${\bf Z}_3$ model with the projection
onto the ${\bf Z}_2$ invariant states (note that 
${\bf Z}_6\approx{\bf Z}_3\otimes {\bf Z}_2$). In particular, the 9 
fixed points in the original ${\bf Z}_3$ twisted sector are combined into 5
linear combinations invariant under ${\bf Z}_2$ plus 4 linear combinations 
which pick up a minus sign under the ${\bf Z}_2$ action. Taking this into 
account, it is straightforward to determine the ${\bf Z}_2$ invariant states
in the ${\bf Z}_3$ twisted 99 open string sector (see \cite{NP} for details).
The ${\bf Z}_3$ twisted 55 open string sector is a bit more subtle - here we
assume that all D5-branes are sitting on top of each other at the orientifold
5-plane located at the origin of K3 ($z_1=z_2=0$). This implies that the
D5-branes only feel the singularity in K3 located at the origin. That is,
the other 8 of the original 9 fixed points in the ${\bf Z}_3$ twisted sector
play no role in this discussion as the twisted 55 states arise due to the {\em 
local} geometry near the origin. This results in the fact that the 
corresponding multiplicity in the ${\bf Z}_3$ twisted 55 open string 
sectors is 1. Finally, the ``$\Omega g$- and $\Omega g^5$-twisted'' sectors 
(with $g^6=1$ in this case) correspond to the ${\bf Z}_3$ twisted 59 open 
string sector. In this case local geometry once again determines the 
multiplicity of states to be 1 (and also fixes their quantum numbers) 
\cite{NP}. The massless spectrum of the ${\bf Z}_6$ model is summarized in
Table I. Note that the twisted 99 and 55 open string sectors no longer
exhibit the naive ``T-duality''. This is due to the fact that these sectors 
do not arise via a straightforward orbifold reduction of the corresponding 
($SO(32)$) gauge theory (with ${\cal N}=2$ supersymmetry in six dimensions).
More concretely, the T-duality transformation involves ${\bf Z}_2$ reflection 
of the complex coordinates $z_1,z_2$. The ${\bf Z}_3$ singularities, however,
are {\em not} invariant under this ${\bf Z}_2$ action, hence the lack of
T-duality symmetry.
      
{}Finally, let us consider the ${\bf Z}_4$ case which was not discussed in 
\cite{NP}. The main difficulty in this case is that (unlike in the ${\bf Z}_3$ 
and ${\bf Z}_6$ cases) this model has no perturbative heterotic dual, nor
do we have any sectors of the theory which can be understood by performing
a perturbative computation on the heterotic side. Fortunately, however, anomaly
cancellation requirements (which are quite constraining in six dimensions)
along with the relevant geometric picture allow us to determine the massless 
spectrum in this case as well. Thus, the gravitational anomaly cancellation
condition reads:
\begin{equation}\label{anom}
 n_H-n_V=273-29n_T~,
\end{equation}
where $n_H,n_V,n_T$ are the numbers of hypermultiplets, vector multiplets
and tensor multiplets, respectively. In this case we have $n_T=1$ tensor 
multiplets and $n^c_H=20$ hypermultiplets from the closed string sector.
All the vector multiplets arise in the open string sector. In particular,
$n_V=256$ as the gauge group is $[U(8)\otimes U(8)]_{99}\otimes
[U(8)\otimes U(8)]_{55}$. The anomaly cancellation condition (\ref{anom}) then
implies that the total number of open string hypermultiplets must be: $n^o_H=
n_H-n^c_H=480$. The perturbative sectors, namely, the untwisted 99 plus 55 as 
well as 59 sectors give rise to 368 hypermultiplets in this model (see 
Table I). Thus,
$112=4\times 28$ hypermultiplets must come from the twisted open string 
sectors. These are the ${\bf Z}_4$ twisted 99 and 55 open string sectors
corresponding to the ``$\Omega g$- and $\Omega g^3$-twisted'' sectors (with
$g^4=1$ in this case). Note that there is no ``twisted'' 59 sector in this 
model. 

{}From the anomaly cancellation we get a hint that the twisted 99 and 55 open
string sector hypermultiplets must transform in the antisymmetric 
representation ${\bf 28}$ of the $SU(8)$ subgroups. In fact, since the two
$SU(8)$'s within, say, the 55 open string sector are on the equal footing,
we conclude that the only way we can cancel anomalies is by assuming that
in the twisted 
55 sector we have one copy of twisted hypermultiplets in $({\bf 28},
{\bf 1})_{55}$ and $({\bf 1},{\bf 28})_{55}$, and similarly for 
the twisted 99 sector.
In the 55 sector this is actually what one expects from the geometric picture:
just as in the ${\bf Z}_6$ case here we have all D5-branes on top of each other
at the orientifold 5-plane located at the origin of K3, so we expect only {\em
one} copy of non-perturbative twisted hypermultiplets arising due to the local
geometry at the vicinity of the origin (the latter being the relevant 
singularity in K3). However, for the 99 sector the conclusion that we have only
one copy of twisted hypermultiplets might appear a bit puzzling - the D9-branes
wrap the entire K3, and naively one expects that they should ``feel'' all 4
${\bf Z}_4$ fixed points. 

{}Let us try to understand this point a bit better. An important observation 
here is that the naive T-duality {\em is} expected to be a symmetry of this
background - indeed, the ${\bf Z}_2$ reflection of the complex coordinates
$z_1,z_2$ involved in the T-duality transformation leaves the corresponding 
${\bf Z}_4$ singularities invariant. Thus, the ${\bf Z}_4$ 
twisted 55 and 99 open string sectors should have T-duality symmetry. Then we
must explain how come in the 99 sector we do not get 4 copies of the 
corresponding matter representations but only one. The point here is that we 
expect massless non-perturbative states to arise only for true geometric 
singularities. However, if there is a non-zero twisted $B$-field turned on
inside of the ${\bf P}^1$ associated with a given blow-up mode (that is, with
a given fixed point), then non-perturbative states do not arise \cite{aspin}.
Since we are dealing with the ${\bf Z}_4$ twisted sectors, the corresponding 
twisted $B$-field can only take values $0,\pm 1/4,1/2$ (here we normalize the
$B$-field so that it is defined modulo a unit shift). The $B$-filed must be
zero at the origin. At the other three fixed points, however, it must be 
non-zero. Note that it cannot be $\pm 1/4$ as this would not be invariant 
under the orientifold action $\Omega$ (recall that the $B$-field is odd under
the action of $\Omega$). However, the $B$-field at these three fixed points can
take a half-integer value which (taking into account that it is only defined 
modulo an integer shift) is invariant under the action of $\Omega$. Thus, to 
obtain a consistent ${\bf Z}_4$ orbifold background (with all D5-branes 
located at the origin of K3), we must turn on half-integer twisted $B$-field
at the other three fixed points. Then the massless spectrum of the ${\bf Z}_4$
model, which is summarized in Table I, is free of anomalies. 
Here we should note that in the 
${\bf Z}_3$ and ${\bf Z}_6$
models the twisted $B$-field (in the ${\bf Z}_3$ twisted sectors) must be zero
as non-zero values such as $\pm 1/3$ would not be invariant under the action of
$\Omega$. This is why in these models all singularities (that is, fixed points)
are relevant in the twisted 99 sector.

{}Here we would like to 
point out that all three ${\bf Z}_2,{\bf Z}_4,{\bf Z}_6$
models are on the same moduli. As we have already mentioned, at generic points
with completely broken gauge symmetry these models correspond to F-theory
compactifications on the Calabi-Yau three-fold (with
the Hodge numbers $(h^{1,1},h^{2,1})=(3,243)$) given by the $T^2$ fibration 
over the base ${\bf P}^1\otimes {\bf P}^1$. 

\section{Type I on Voisin-Borcea Orbifolds}

{}In this section we consider four dimensional ${\cal N}=1$ supersymmetric
Type I compactifications on Calabi-Yau three-folds known as Voisin-Borcea
orbifolds \cite{Voisin,Borcea}. 
To aid the presentation, in the next subsection we review various 
basic facts about Voisin-Borcea orbifolds following \cite{Asp,MV,KST}. To 
clarify some points discussed in the previous sections, we also briefly discuss
F-theory compactifications on these spaces. Having reviewed the relevant
background material, we then discuss Type I compactifications on these 
Calabi-Yau three-folds as well as their dual F-theory compactifications on 
the corresponding Calabi-Yau four-folds known as Borcea four-folds.

\subsection{Voisin-Borcea Orbifolds}

{}Let ${\cal W}_2$ be a K3 surface (which is not necessarily an 
orbifold of $T^4$) which admits an involution $J$ such that it reverses the 
sign of the holomorphic
two-form $dz_1\wedge dz_2$ on ${\cal W}_2$. Consider the following quotient:
\begin{equation}
 {\cal Y}_3= (T^2\otimes {\cal W}_2)/Y~,
\end{equation}
where $Y=\{1,S\}\approx {\bf Z}_2$, and $S$ acts as $Sz_0=-z_0$ on $T^2$ 
($z_0$ being a complex coordinate on $T^2$), and as $J$ on ${\cal W}_2$. This
quotient is a Calabi-Yau three-fold with $SU(3)$ holonomy 
which is elliptically 
fibered over the base ${\cal B}_2={\cal W}_2/B$, where 
$B=\{1,J\}\approx{\bf Z}_2$. 

{}Nikulin gave a classification \cite{Nik} 
of possible involutions of K3 surfaces in terms of three 
invariants $(r,a,\delta)$
(for a physicist's discussion, see, {\em e.g.}, \cite{Asp,MV}).
The result of this classification is plotted in Fig.2 (which has been borrowed
from \cite{KST})
according to the values of $r$ 
and $a$. The open and closed circles correspond to the cases with $\delta=0$
and $\delta=1$, respectively. (The cases denoted by ``$\otimes$'' are outside
of Nikulin's classification, and we will discuss them shortly.) In the case 
$(r,a,\delta)=(10,10,0)$ the base ${\cal B}_2$ is an Enriques surface, and the 
corresponding ${\cal Y}_3$ has Hodge numbers $(h^{1,1},h^{2,1})=(11,11)$.
In all the other cases the Hodge numbers are given by:
\begin{eqnarray}\label{hodge1}
 &&h^{1,1}=5+3r-2a~,\\
 \label{hodge2}
 &&h^{2,1}=65-3r-2a~.
\end{eqnarray} 

{}For $(r,a,\delta)=(10,10,0)$ the ${\bf Z}_2$ twist $S$ is freely acting 
(that is, it has no fixed points). 
For $(r,a,\delta)=(10,8,0)$ the fixed point set
of $S$ consists of two curves of genus 1. The base ${\cal B}_2$ in this case 
is ${\bf P}^2$ blown up at 9 points. In all the other cases the fixed point set
of $S$ consists of one curve of genus $g$ plus $k$ rational curves, where
\begin{eqnarray}
 &&g={1\over 2}(22-r-a)~,\\
 &&k={1\over 2}(r-a)~.
\end{eqnarray}

{}It is sometimes useful to separate the above Hodge numbers into the
contribution $(h^{1,1}_0,h^{2,1}_0)$ of the untwisted sector and the
contribution $(h^{1,1}_*,h^{2,1}_*)$ of the ${\bf Z}_2$ twisted (that is,
$S$-twisted) sector. These are given by
\begin{eqnarray}\label{hodge10}
 &&h^{1,1}_0=r+1~,\\
 \label{hodge20}
 &&h^{2,1}_0=21-r~,
\end{eqnarray}
and
\begin{eqnarray}\label{hodge1*}
 &&h^{1,1}_*=4(k+1)~,\\
 \label{hodge2*}
 &&h^{2,1}_*=4g~,
\end{eqnarray}  
respectively.

{}Note that except for the cases with $a=22-r$, 
$r=11,\dots,20$, the mirror pair
of ${\cal Y}_3$ is given by the Voisin-Borcea orbifold 
${\widetilde {\cal Y}}_3$ with ${\widetilde r}=20-r$, ${\widetilde a}=a$.
Under the mirror transform we have: ${\widetilde g}=f$, ${\widetilde f}=g$, 
where $f=k+1$. 

{}In the cases $a=22-r$, $r=11,\dots,20$, the mirror would have to have
${\widetilde r}=20-r$ and ${\widetilde a}=a={\widetilde r}+2$, where 
${\widetilde r}=0,\dots,9$. We have depicted these cases in Fig.2 using 
the ``$\otimes$'' symbol. In 
particular, 
we have plotted cases with $a=r+2$, $r=0,\dots,10$.
The Hodge numbers for these 
cases are still given by (\ref{hodge1}) and (\ref{hodge2}) 
(which follows from
their definition as mirror pairs of the cases with $a=22-r$, 
$r=11,\dots,20$). (This is
true for $a=r+2$, $r=0,\dots,9$. 
Extrapolation to $r=10$ is motivated by 
the fact that
in this case we get $(h^{1,1},h^{2,1})=(11,11)$ which is the same as for 
$(r,a,\delta)=(10,10,0)$.) In \cite{KST} it was argued that these Voisin-Borcea
orbifolds also exist, albeit they are {\em singular}. 
In fact, some of them can be constructed explicitly (see
\cite{KST} for details).

{}As we already mentioned, here we would like to briefly review
F-theory compactifications on Voisin-Borcea orbifolds (which correspond to
${\cal N}=1$ supersymmetric vacua in six dimensions). Thus, consider F-theory
on ${\cal Y}_3$ with\footnote{In the case $(r,a,\delta)=(10,10,0)$ 
we have $T=9$ tensor multiplets, $H=12$ hypermultiplets, and no gauge 
vector multiplets. In the case $(r,a,\delta)=(10,8,0)$ 
we have $T=9$ tensor multiplets, $H=12$ neutral hypermultiplets,
gauge vector multiplets corresponding to $SO(8)\otimes SO(8)$, one 
hypermultiplet in the adjoint of the first $SO(8)$, and one hypermultiplet in 
the adjoint of the second $SO(8)$.} $(r,a,\delta)\not=
(10,10,0)$ or $(10,8,0)$. This
gives rise to the following massless spectrum in six 
dimensions. The number of tensor multiplets is $T=r-1$. The number of neutral 
hypermultiplets is $H=22-r$. The gauge group is $SO(8)\otimes SO(8)^k$.
There are $g$ adjoint hypermultiplets of the first $SO(8)$. There are no 
hypermultiplets charged under the other $k$ $SO(8)$'s. Under mirror symmetry
$g$ and $f=k+1$ are interchanged. Thus, the vector multiplets in the adjoint
of $SO(8)^k$ are traded for $g-1$ hypermultiplets in the adjoint of the first 
$SO(8)$. That is, gauge symmetry turns 
into global symmetry and {\em vice-versa}.
This can be pushed further to understand F-theory compactifications on
Calabi-Yau three-folds 
with $a=r+2$, $r=1,\dots,10$, which give the following 
spectra\footnote{Here we 
must exclude the case with $r=0$, $a=2$ for the F-theory prediction 
would be $T=-1$
tensor multiplets. This Calabi-Yau three-fold 
does exist \cite{KST}, 
but it is singular and F-theory compactification on this space does 
not appear to
have a local Lagrangian description. However, 
an extremal transition \cite{MV}
between this Calabi-Yau 
three-fold and another Voisin-Borcea 
orbifold could lead to a phase transition into
a well defined vacuum.}. The number of tensor multiplets is $T=r-1$. There are 
$H=22-r$ neutral hypermultiplets. In 
addition there are $g=10-r$ hypermultiplets
transforming as adjoints under a global $SO(8)$ symmetry. 
There are no gauge bosons,
however. It is not difficult to verify that this massless 
spectrum is free of gravitational
anomalies in six dimensions. In fact, F-theory compactifications
on these singular spaces are equivalent to F-theory compactifications
on smooth Calabi-Yau three-folds according to the following relation 
\cite{KST}:
\begin{eqnarray}
 &&{\mbox{F-theory on 
 ${\cal Y}_3$ with $(h^{1,1},h^{2,1})=(r+1,61-5r)$ is equivalent
 to}}\nonumber\\
 &&{\mbox{F-theory on 
 ${\widehat{\cal Y}}_3$ with $({\hat h}^{1,1},{\hat h}^{2,1})
 =(r+1,301-29r)$  ($r=1,\dots,9$)}}~.
\end{eqnarray}
Thus, for instance, 
for $r=2$ we get $({\hat h}^{1,1},{\hat h}^{2,1})=(3,243)$, 
which is the elliptic Calabi-Yau given by the $T^2$ fibration over
the base ${\bf P}^1\otimes {\bf P}^1$. We have already 
encountered this Calabi-Yau three-fold in the previous sections. Note that
F-theory compactifications on Voisin-Borcea orbifolds with $(r,a,\delta)=
(2,2,0),(2,2,1)$ are on the same moduli as the background corresponding
to $({\hat h}^{1,1},{\hat h}^{2,1})=(3,243)$. The reason is that the
corresponding bases in the former two cases (namely, the Hirzebruch surfaces
${\bf F}_0$ and ${\bf F}_1$) are singular at the Voisin-Borcea points
\cite{MV}.
After the appropriate blow-ups additional matter arises precisely such that
the moduli count matches that for the $({\hat h}^{1,1},{\hat h}^{2,1})=
(3,243)$ compactification. 
  
\subsection{Type I on Voisin-Borcea Orbifolds}

{}In this subsection we consider some general aspects of Type I 
compactifications on Voisin-Borcea orbifolds reviewed in the previous 
subsection. In discussing such a compactification, we must specify the gauge
bundle. We discussed perturbative (from the orientifold viewpoint) 
gauge bundles for K3 compactifications in section II. Our goal here will be
to determine which gauge bundles are perturbative from the orientifold 
viewpoint for the Voisin-Borcea compactifications. It turns out that one can
answer this question using the map between Type IIB orientifolds and F-theory.
In fact, as we will see in a moment, the requirement that the gauge bundle
be perturbative greatly 
restricts the possible choices of the compactification space itself.

{}Thus, consider Type I compactification on the Voisin-Borcea orbifold
${\cal Y}_3$ labeled by $(r,a,\delta)\not=(10,10,0)$. We can view this
as the $\Omega$ orientifold of Type IIB on ${\cal Y}_3$. This theory contains
32 D9-branes as well as D5-branes. Thus, there are D5-branes filling 
${\bf R}^{3,1}\otimes {\cal C}$, where ${\bf R}^{3,1}$ is the non-compact four
dimensional Minkowski space-time, whereas ${\cal C}$ is the set of points
(of real dimension 2) in ${\cal B}_2$ fixed under the action of the 
${\bf Z}_2$ twist $S$ (see the previous subsection for notations). There may be
other 5-branes depending upon the choice of the space ${\cal Y}_3$. Thus,
if the K3 surface ${\cal W}_2$ (recall that the base of the elliptic fibration
is ${\cal B}_2={\cal W}_2/J$) can be written as the quotient ${\cal W}_2=
{\cal W}_2^\prime/{\bf Z}_2$ (here ${\cal W}^\prime_2$ can be either $T^4$ or 
another K3 surface), where the generator $R$ of ${\bf Z}_2$ reflects both of 
the complex coordinates $z_1,z_2$ on ${\cal W}^\prime_2$, then we have other 
5-branes in the theory. First, we have D5-branes filling ${\bf R}^{3,1}\otimes
{\cal C}^\prime$, where ${\cal C}^\prime$ is the set of points (of real 
dimension 2) in ${\cal B}_2$ fixed under the action of the twist $SR$. Also,
there are 5-branes filling ${\bf R}^{3,1}\otimes T^2$ (here $T^2$ is the fibre
$T^2$). More precisely, as we will see in a moment, 
this is the case for perturbative gauge bundles (while
for other choices of the gauge bundle these 5-branes may be absent).

{}To understand which gauge bundles are perturbative from the orientifold 
viewpoint in the case of Voisin-Borcea compactifications, it is convenient
to T-dualize the corresponding Type I background in the direction of 
the fibre $T^2$. This way we arrive at the $\Omega {\widetilde J}(-1)^{F_L}$
orientifold of Type IIB on ${\cal Y}_3$ (which contains various
7-branes plus, in certain cases, 3-branes). Here ${\widetilde J}$ 
reflects the
complex coordinate $z_0$ on the fibre $T^2$ (while acting trivially on the base
${\cal B}_2$), and $F_L$ is the left-moving
(space-time) fermion number. This orientifold can now be mapped to the 
F-theory compactification on an elliptically fibered Calabi-Yau four-fold
via the map of \cite{sen} which we review next. 

{}Consider an $\Omega {\widetilde J} (-1)^{F_L}$ orientifold of Type IIB on 
${\cal Y}_3$, where ${\widetilde J}$ 
reverses the sign of the holomorphic 3-form 
$dz_1\wedge dz_2\wedge dz_3$ on ${\cal Y}_3$, and the 
set of points fixed under ${\widetilde J}$ in ${\cal Y}_3$ has real 
dimension 4. Then following \cite{sen}
we can map this orientifold to (a limit of) F-theory on a 
Calabi-Yau four-fold ${\cal X}_4$ defined as
\begin{equation}\label{CY_4}
 {\cal X}_4=({\widetilde T}^2\otimes {\cal Y}_3)/X~,
\end{equation}     
where $X=\{1,{\widetilde S}\}\approx{\bf Z}_2$, and ${\widetilde S}$ acts as 
${\widetilde S} {\widetilde z}_0=-{\widetilde z}_0$ on ${\widetilde T}^2$ 
(${\widetilde z}_0$ is a complex coordinate on ${\widetilde T}^2$), 
and as ${\widetilde J}$ 
on ${\cal Y}_3$. To avoid confusion, here we use 
tilde to distinguish between the
fibre ${\widetilde T}^2$ in the elliptic fibration over the base  
${\cal B}_3={\cal Y}_3/B$, where $B\equiv\{1,
{\widetilde J}\}\approx{\bf Z}_2$, and the fibre $T^2$ in the elliptic 
fibration over the base ${\cal B}_2={\cal W}_2/Y$.

{}Note that in the case where ${\cal Y}_3$ is a Voisin-Borcea orbifold, the
above four-fold ${\cal X}_4$ can be viewed as a quotient $({\widetilde 
{\cal W}}_2 \otimes {\cal W}_2)/Y$. Here ${\widetilde {\cal W}}_2\equiv
({\widetilde T}^2\otimes T^2)/X$ (recall that 
$X=\{1,{\widetilde S}\}\approx{\bf Z}_2$, and ${\widetilde S}$ acts as
${\widetilde S}{\widetilde z}_0=-{\widetilde z}_0$ and ${\widetilde S}
z_0=-z_0$ on the complex coordinates parametrizing ${\widetilde T}^2$ and 
$T^2$, respectively). Thus, ${\widetilde{\cal W}}_2$ is (the ${\bf Z}_2$ 
orbifold limit of) a K3 surface. The action of $Y=\{1,S\}$ on the two
K3 surfaces ${\widetilde {\cal W}}_2$ and ${\cal W}_2$ is given by
the corresponding Nikulin's involutions labeled by the integers
$({\widetilde r},{\widetilde a},{\widetilde \delta})$ and $(r,a,\delta)$
respectively. (In fact, it is not difficult to show that in the present case 
${\widetilde r}=18$ and ${\widetilde a}=4$.) These Calabi-Yau four-folds 
(which have $SU(4)$ holonomy) generically are singular spaces. The Euler 
number $\chi$ of such a four-fold is given by \cite{Borcea}:
\begin{equation}
 {1\over 24} \chi= 12+{1\over 4}({\widetilde r} -10) (r-10)~.
\end{equation}   
Taking into account that in the case at hand we have ${\widetilde r}=18$,
we obtain
\begin{equation}
 {1\over 24}\chi=12+2(r-10)=2(r-4)~.
\end{equation}
Now consider F-theory compactification on such a four-fold. To cancel
space-time anomaly we need to introduce $\chi/24$ 3-branes, which implies that
this number must be non-negative or else supersymmetry appears to be broken
\cite{SVW}. It then follows that
\begin{equation}\label{state}
 {\mbox{Type I on Voisin-Borcea orbifolds can be tadpole free only for
 $r\geq 4$.}}
\end{equation}
This is a non-trivial constraint with interesting implications for Type I
model building. For instance, if we take Type I on a ${\bf Z}_2\otimes 
{\bf Z}_2$ orbifold of $T^2\otimes T^2\otimes T^2$ with discrete torsion, the
Hodge numbers are $(h^{1,1},h^{2,1})=(3,51)$ (note that this is the mirror
Calabi-Yau of a usual ${\bf Z}_2\otimes {\bf Z}_2$ orbifold without discrete
torsion whose Hodge numbers are $(h^{1,1},h^{2,1})=(51,3)$), and we have
$r=2$ and $a=4$. Thus, Type I compactification on this space is always
anomalous (that is, there is no choice of the gauge bundle consistent with
tadpole cancellation). 

{}Thus, let us consider Type I compactifications with $r\geq 4$. In the dual
F-theory picture the number of 3-branes required for anomaly cancellation
is $2(r-4)$. If this number is non-zero, we have a choice of where to place
these 3-branes: ({\em i}) we can keep them in the bulk; from the Type I
point of view these correspond to dynamical 5-branes (made of some number of 
D5-branes); ({\em ii}) 
alternatively, we can ``dissolve'' them into the 7-branes;
from the Type I viewpoint this corresponds to embedding a certain
${\mbox{Spin}}(32)/{\bf Z}_2$ gauge bundle. The corresponding instantons are no
longer point-like (at generic points). Thus, we see that we need to specify 
additional data, which is 
the choice of the gauge bundle. What we are interested
in understanding here is which choices are perturbative from the orientifold
viewpoint. 

{}To answer this question recall that in K3 compactifications the total
number of instantons must be 24. By these we actually mean the number 
$n_I$ of instantons embedded in the gauge bundle plus the number $n_5$ of
5-branes transverse to K3. In the case of the Voisin-Borcea compactifications
this number 24 is reduced to 12 - the corresponding pairing takes place
due to the additional ${\bf Z}_2$ projection by the orbifold group $Y$. Thus,
$\chi/24$ must be at least 12. This implies that the statement (\ref{state})
can be made even stronger: 
\begin{equation}\label{state1}
 {\mbox{Type I on Voisin-Borcea orbifolds can be tadpole free only for
 $r\geq 10$.}}
\end{equation}     
In fact, since we have a ${\bf Z}_2$ twist generated by $S$, we must include
32 D5-branes (filling ${\bf R}^{3,1}\otimes {\cal C}$ - see above). 
These correspond to 8 dynamical 5-branes: here two pairings take place - one
due to the $\Omega$ orientifold projection, and the other one due to the ${\bf
Z}_2$ orbifold projection. Note that here we are making an assumption that
the corresponding background is perturbative from the orientifold viewpoint -
it is {\em a priori} not obvious at all that in other cases we must 
include D5-branes. In fact, here we can give an example where this is {\em not}
the case. Thus, consider ${\mbox{Spin}}(32)/{\bf Z}_2$ 
heterotic on $T^4/{\bf Z}_2$ with the standard 
embedding of the gauge bundle (which we discussed in section II). This theory
is perturbative from the heterotic viewpoint, hence no 5-branes are present.
That is, no D5-branes would be present on the dual Type I side either. This
compactification, however, is {\em non-perturbative} from the orientifold 
viewpoint.

{}Thus, we must have at least 20 3-branes (12 from K3, and 8 from the D5-branes
filling ${\bf R}^{3,1}\otimes {\cal C}$). This implies that, for the gauge
bundle to be perturbative, we must have $r\geq 14$. In fact we will now show 
that there are only two choices of $r$ for which we can have perturbative
gauge bundle: $r=14$ and $r=18$. First, suppose $r>14$. Then we must have
additional 5-branes whose number is $2(r-14)$. However, the only other way
we can obtain 5-branes perturbatively in the orientifold language is to
have 8 dynamical 5-branes (arising from 32 D5-branes) filling
${\bf R}^{3,1}\otimes {\cal C}^\prime$ (these are perpendicular to those
filling ${\bf R}^{3,1}\otimes {\cal C}$). This then implies that we must have
$r=18$. On the other hand, if $r=14$ we must make sure that there are no
5-branes filling ${\bf R}^{3,1}\otimes {\cal C}^\prime$. This implies that
in this case the K3 surface ${\cal W}_2$ {\em cannot} be written as the 
quotient ${\cal W}^\prime/{\bf Z}_2$. It then follows that there are {\em no}
5-branes transverse to K3 (that is, ${\cal W}_2$) either (that is, $n_5=0$ and
$n_I=24$ in the corresponding K3 compactification).      

{}The above discussion shows that the gauge bundle can be perturbative if and
only if $r=14$ or $r=18$. Let us apply these constraints to the cases where
the K3 surface ${\cal W}_2$ is a toroidal orbifold, that is, ${\cal W}_2=
T^4/{\bf Z}_N$ with $N=2,3,4,6$. In the $N=2$ case (without discrete torsion 
between $S$ and the generator $g$ of the K3 orbifold group) the Hodge numbers
of the corresponding Voisin-Borcea orbifold are $(h^{1,1},h^{2,1})=(51,3)$, and
$r=18$, $a=4$. We then arrive at a consistent Type I compactification on
the ${\bf Z}_2\otimes {\bf Z}_2$ orbifold discussed in \cite{BL}. In the $N=3$
case the Hodge numbers are $(h^{1,1},h^{2,1})=(35,11)$, and we arrive at the
non-perturbative orientifold corresponding to Type on ${\bf Z}_6^\prime$ 
orbifold recently constructed in \cite{NP}. For later convenience we give 
the massless spectrum of this model in Table II.  

{}Next, consider the $N=4$ case. Suppose there is no discrete torsion between
$S$ and $g^2$ (where $g$ is the generator of the K3 orbifold group). Then the
Hodge numbers are $(h^{1,1},h^{2,1})=(61,1)$, and, therefore, $r=20$ and $a=2$.
On the other hand, if we include discrete torsion between $S$ and $g^2$, we 
obtain $(h^{1,1},h^{2,1})=(3,51)$ just as in the ${\bf Z}_2\otimes {\bf Z}_2$ 
case with discrete torsion. Thus, in Type on the ${\bf Z}_2\otimes {\bf Z}_4$ 
orbifold we cannot cancel tadpoles if we choose the perturbative gauge bundle.
(In fact, in the case with
discrete torsion it is impossible to cancel tadpoles for any choice of the
gauge bundle. In the case without discrete torsion tadpole
cancellation might be possible
at the expense of choosing a gauge bundle which is non-perturbative from the
orientifold viewpoint.) This explains why the naive perturbative tadpole 
cancellation conditions were found in \cite{Zw} to have no solution.
Here the map to F-theory provides a simple geometric interpretation of this 
fact.

{}The above conclusion about perturbative inconsistency of Type I on
${\bf Z}_2\otimes {\bf Z}_4$ can be used to understand other similar
models. Note that the sectors of the orientifold labeled by $\Omega$,
$\Omega S$, $\Omega g^k$, $k=1,2,3$, and $\Omega Sg^2$ are perturbatively well
defined (as the corresponding K3 orientifolds are well defined). This implies
that the ``troublesome'' sectors are the $\Omega Sg$ and $\Omega Sg^3$ sectors.
This is precisely the conclusion that was reached in \cite{KST}. 
Note that $Sg$ is the generator of an orbifold group ${\bf Z}_4$, which, to
avoid confusion, we will refer to as ${\bf Z}_4^*$. In particular, 
$T^6/{\bf Z}_4^*$ is a Calabi-Yau three-fold with $SU(3)$ holonomy. Thus, 
we conclude that Type I on the ${\bf Z}_4^*$ orbifold is not consistent for
perturbative gauge bundles. In fact,
we can further deduce that perturbative gauge bundles would lead to
inconsistencies in the ${\bf Z}_4\otimes {\bf Z}_4$, ${\bf Z}_8$ and 
${\bf Z}_{12}^\prime$ orbifold models as well since all of these orbifold 
groups contain ${\bf Z}_4^*$ as a subgroup. This is in accord with the 
discussion in \cite{KST}, where it was also argued why in the 
${\bf Z}_8^\prime$ model choosing the perturbative gauge bundle leads to
similar inconsistencies as well. Thus, the discussion in \cite{KST} explains 
the results of \cite{ibanez}
(where it was shown that the naive perturbative tadpole cancellation conditions
in the ${\bf Z}_4^*$, ${\bf Z}_8$, ${\bf Z}_8^\prime$ and ${\bf Z}_{12}^\prime$
cases have no solution) using the geometric picture via
the map to F-theory. Here the ${\bf Z}_2\otimes {\bf Z}_4$ example serves
as an illustration of the discussion in \cite{KST}.  

{}Finally, let us discuss the case $N=6$ (without discrete torsion between 
$S$ and $g^3$, where $g$ is the generator of the K3 orbifold group). The
Hodge numbers in this case are $(h^{1,1},h^{2,1})=(51,3)$ just as in the 
${\bf Z}_2\otimes {\bf Z}_2$ case without discrete torsion. This implies that
$r=18$ and $a=4$. This compactification, therefore, should be consistent
for the appropriate perturbative gauge bundle. We will
explicitly construct this model (via the corresponding non-perturbative 
orientifold) in the next subsection. 

\subsection{The ${\bf Z}_2\otimes {\bf Z}_6$ Model}

{}In this subsection we are going to give the massless spectrum of the
non-perturbative orientifold based on the ${\bf Z}_2\otimes {\bf Z}_6$ orbifold
compactification of Type I theory. This background has ${\cal N}=1$ 
supersymmetry in four dimensions. Here we should emphasize that the choice of
the gauge bundle in the model we discuss in this subsection is 
{\em perturbative} from the orientifold viewpoint. The non-perturbative sectors
arise due to singularities in the Calabi-Yau itself as discussed in 
Introduction. That is, non-perturbative states in this theory arise along the
lines of our discussion in section III for K3 compactifications of Type I.

{}Thus, consider Type I on the Voisin-Borcea orbifold ${\cal Y}_3=
(T^2\otimes T^2\otimes T^2)/({\bf Z}_2\otimes {\bf Z}_6)$, where
the generators $R_3$ and $g$ of ${\bf Z}_2$ respectively ${\bf Z}_6$ have
the following action on the complex coordinates $z_1,z_2,z_3$ parametrizing the
three two-tori:
\begin{eqnarray}
 &&R_iz_j=-(-1)^{\delta_{ij}}z_j~,~i,j=1,2,3~,\\
 &&\theta z_1=z_1~,~~~\theta z_2=\alpha z_2~,~~~\theta z_3=\alpha^{-1} z_3~,
\end{eqnarray}
where $R_1\equiv g^3$, $R_2\equiv R_1 R_3$, $\theta\equiv g^2$, and
$\alpha\equiv\exp(2\pi i/3)$. The Calabi-Yau three-fold ${\cal Y}_3$ 
(which has $SU(3)$ holonomy) has the Hodge numbers $(h^{1,1},h^{2,1})=
(51,3)$, so that there are $54$ chiral supermultiplets in the closed string 
sector.

{}In the open string sector we have 32 D9-branes, and also three sets of 
D5-branes with 32 D5-branes in each set. Thus, the world-volumes of the 
D$5_i$-branes are four non-compact space-time dimensions plus the two-torus
parametrized by the complex coordinate $z_i$. For the sake of definiteness,
in the following we will concentrate on the brane configuration where all 
D$5_i$-branes are placed on top of each other at the orientifold $5_i$-plane
located at the origin in the transverse dimensions (that is, $z_j=0$, 
$j\not=i$). 

{}Up to equivalent representations the Chan-Paton matrices are given by:
\begin{eqnarray}
 &&\gamma_{\theta,9}={\mbox{diag}}({\bf W}\otimes {\bf I}_2,{\bf I}_8)~,\\
 &&\gamma_{R_i,9}=i\sigma_i\otimes {\bf I}_8~.
\end{eqnarray}
Here ${\bf W}\equiv{\mbox{diag}}(\alpha,\alpha,\alpha^{-1},\alpha^{-1})$,
and $\sigma_i$ are the usual $2\times 2$ Pauli matrices. (The action on
the D$5_i$-branes is similar.) The perturbative (from the orientifold 
viewpoint) massless open string spectrum of this model can be obtained
using the standard orientifold techniques, and is 
given\footnote{Here we note that the perturbative spectrum of this model
was discussed in \cite{Zw}.} 
in Table III.
Thus, the $\Omega$-twisted sector corresponds to the untwisted 99 plus 
$5_i5_i$ open string sectors, while $\Omega R_i$ twisted sector
corresponds to the untwisted $95_i$ plus $5_j5_k$ ($j\not=k\not=i$) open string
sectors.

{}The non-perturbative sectors in this orientifold are the following.
The $\Omega\theta$- and $\Omega\theta^2$-twisted sectors
correspond to the twisted 99 plus $5_i5_i$ open string sectors, while
the $\Omega R_i\theta$- and $\Omega R_i\theta^2$-twisted sectors
correspond to the twisted $95_i$ plus $5_j5_k$ ($j\not=k\not=i$)
open string sectors. The twisted $95_1$ and $95_{2,3}$ sectors 
are straightforward to work out
starting from the twisted 95 sector in the six dimensional 
${\bf Z}_6$ model in Table I respectively the twisted 95 sector in the 
four dimensional ${\bf Z}_6^\prime$ model in Table II and projecting onto the
corresponding ${\bf Z}_2$ invariant states. The twisted $5_25_3$ sector is 
the same as the twisted $95_1$ sector (with the obvious substitutions of the
gauge quantum numbers). The twisted $5_1 5_{2,3}$ sectors are the same as 
the twisted $9 5_{3,2}$ sectors except that the multiplicity of the states in
the latter is 1, while it is 3 in the former. This is due to the fact that 
the the corresponding 5-branes only feel the local geometry in the vicinity
of the corresponding fixed points at which they are placed. (This is in 
complete parallel with the corresponding discussion for the six dimensional
cases in \cite{NP} and section III.) As to the twisted $99$, $5_2 5_2$ and 
$5_3 5_3$ sectors, they can be worked out by starting from the twisted
99 and 55 sectors in the ${\bf Z}_6^\prime$ model in Table II and projecting
onto the corresponding ${\bf Z}_2$ invariant states. Similarly, the twisted 
$5_1 5_1$ sector (as well as the twisted 99 sector) can be worked out by
starting from the twisted 55 sector in the ${\bf Z}_6$ model in Table I and
projecting onto the corresponding ${\bf Z}_2$ invariant states.
The complete massless spectrum of the ${\bf Z}_2\otimes {\bf Z}_6$ model
including both perturbative and non-perturbative (from the orientifold 
viewpoint) states is given in Table III. Note that all non-Abelian gauge
anomalies cancel in this model.

\section{Type I on Generalized Voisin-Borcea Orbifolds}
     
{}In the previous section we considered Type I compactifications on 
Voisin-Borcea orbifolds of the form $(T^2\otimes{\mbox{K3}})/{\bf Z}_2$. Here
we are going to discuss Type I compactifications on generalized Voisin-Borcea
orbifolds of the form 
\begin{equation}
 {\cal Y}_3=(T^2\otimes {\cal W}_2)/{\bf Z}_N~,
\end{equation}
where ${\cal W}_2$ is a K3 surface, and the generator $\eta$ of ${\bf Z}_N$
acts as $\eta z_0=\omega z_0$ on the fibre $T^2$ (parametrized by the complex 
coordinate $z_0$), and as $\eta \Omega_2=\omega^{-1}\Omega_2$ on ${\cal W}_2$
(parametrized by the complex coordinates $z_1,z_2$). Here $\omega\equiv
\exp(2\pi i/N)$, and $\Omega_2\equiv dz_1\wedge dz_2$ is the holomorphic
two-form on ${\cal W}_2$. Note that the holomorphic three-form $\Omega_3\equiv
dz_0\wedge dz_1\wedge dz_2$ on ${\cal Y}_3$ is invariant under the action
of $\eta$, which implies that ${\cal Y}_3$  is a Calabi-Yau three-fold with
$SU(3)$ holonomy (which is actually elliptically fibered over the base
${\cal W}_2/{\bf Z}_N$)
provided that the action of the orbifold group ${\bf Z}_N$ is a symmetry
of $T^2\otimes {\cal W}_2$. The latter requirement implies that $N$ is 
restricted to the values $N=2,3,4,6$ so that the action of ${\bf Z}_N$ on $T^2$
is crystallographic. Note that in the $N=2$ case we have Voisin-Borcea 
orbifolds discussed in the previous section. In this section we will be 
interested in the cases $N=3,4,6$. 

{}Let us start with the $N=4$ case. Suppose ${\cal W}_2$ is an orbifold K3.
Then the three-fold ${\cal Y}_3$ can be either the ${\bf Z}_4\otimes {\bf Z}_2$
or ${\bf Z}_4\otimes {\bf Z}_4$ orbifold of $T^2\otimes T^4$. We have already 
discussed these cases in the previous section where we saw that for 
perturbative gauge bundles tadpoles cannot be canceled in these models.

{}Next, let us consider the $N=6$ case with ${\cal W}_2$ an orbifold K3. Then
${\cal Y}_3$ can be the ${\bf Z}_6 \otimes {\bf Z}_2$, ${\bf Z}_6
\otimes {\bf Z}_3$, ${\bf Z}_6\otimes {\bf Z}_6$ or ${\bf Z}_6^*\otimes 
{\bf Z}_2$ orbifold of 
$T^2\otimes T^4$. The first case was explicitly constructed in the previous
section. We will discuss the last case, which was explicitly constructed
in \cite{223}, in a moment. 
The other two cases are straightforward to work out: the perturbative
part of the spectrum is obtained using the standard orientifold techniques
\cite{Zw}, whereas the non-perturbative twisted open string sector states
can be easily worked out by starting with
the corresponding twisted states in the
${\bf Z}_6$ model in Table I, the ${\bf Z}_6^\prime$ model in Table II, and
the ${\bf Z}_2\otimes {\bf Z}_6$ model in Table III, and performing the 
appropriate (${\bf Z}_2$ and/or ${\bf Z}_3$) orbifold projections. 
Here the following remark is in order. Note that in the 
${\bf Z}_6\otimes {\bf Z}_3$ and ${\bf Z}_6\otimes {\bf Z}_6$ models 
{\em a priori} there are two types of twisted open string sectors. First, we 
have open string sectors corresponding to the orientifold group elements 
labeled by $\Omega\rho$, where $\rho$ is an orbifold group element such that
for some choice of $k\in 2{\bf Z}$ the set of points (in $T^2\otimes 
{\cal W}_2$) fixed under $\rho^k$ is of real dimension 2. These are precisely 
the twisted open string sectors which can be deduced via the appropriate 
orbifold projections of the ``parent'' models mentioned above. This is due to
the fact that such twisted open string sectors are either projections of those
in the corresponding $T^2\otimes{\mbox{K3}}$ compactifications, or of 
twisted 59
open string sectors which can be read off by appropriately projecting that in
the ${\bf Z}_6^\prime$ model in Table II. The second type of twisted open 
string sectors correspond to the orientifold group elements labeled by $\Omega
\rho$ such that for some choice of $k\in 2{\bf Z}$ the set of points fixed 
under $\rho^k$ is zero dimensional. Such twisted open string sectors were 
argued in \cite{KST} to give rise to states that become heavy and decouple
upon appropriately blowing up the corresponding singularities in the
compactification space \cite{ZK,KS,KST}. We will discuss this point in more 
detail shortly. Here, however, an important observation is that the second
type of twisted open string sectors do not give rise to massless states once we
consider {\em blown-up} orbifolds. 
  
{}For the sake of brevity here we will not give the spectra of the
${\bf Z}_6\otimes {\bf Z}_3$ and ${\bf Z}_6\otimes {\bf Z}_6$
models. Let us mention,
however, that the gauge groups in these two cases are 
$[U(2)^6\otimes U(4)]_{99}\otimes [U(2)^6\otimes U(4)]_{55}$ respectively
$[U(2)^3\otimes Sp(4)]_{99}\otimes \bigotimes_{i=1}^3 
[U(2)^3\otimes Sp(4)]_{5_i5_i}$. Thus, the ${\bf Z}_6\otimes {\bf Z}_6$  
model is non-chiral if we ignore the $U(1)$ charges. The 
${\bf Z}_6\otimes {\bf Z}_3$ model is chiral as in the perturbative
open string sectors there are states transforming
under ${\bf 4}$ and ${\overline{\bf 4}}$ of the $SU(4)$ subgroups. However,
the twisted open string sector states are not charged under the $SU(4)$
subgroups, so that these sectors are automatically free of non-Abelian gauge
anomalies. These two models, therefore, are not as illuminating as, say, the
${\bf Z}_2\otimes {\bf Z}_6$ model in Table III.

{}Let us point out that the ${\bf Z}_6^*\otimes {\bf Z}_2$ model was explicitly
constructed in \cite{223}. The discussion of twisted open string sectors in 
this model is completely analogous to that in the ${\bf Z}_6^*(\approx
{\bf Z}_3^*\otimes {\bf Z}_2)$ case. Let us therefore consider the generalized 
Voisin-Borcea orbifolds with $N=3$ (and the ${\bf Z}_6^*$ case is a particular
example of this type).

{}For $N=3$ we have the following possibilities: ${\bf Z}_3^*\otimes {\bf Z}_2
\approx{\bf Z}_6^*$, ${\bf Z}_3^*\otimes {\bf Z}_4\approx{\bf Z}_{12}$,
${\bf Z}_3\otimes {\bf Z}_3$, and ${\bf Z}_3\otimes {\bf Z}_6$. The last case
we have already considered. The ${\bf Z}_3\otimes {\bf Z}_3$ case was 
explicitly
constructed in \cite{KS}. Just as in the ${\bf Z}_6\otimes {\bf Z}_3$ and
${\bf Z}_6\otimes {\bf Z}_3$ cases, here we also have twisted open string 
sectors of two types. As was shown in \cite{KS}, in this model twisted states
of {\em both} types decouple upon appropriately blowing up the orbifold 
singularities, so that the remaining massless spectrum coincides with that 
obtained in the perturbative orientifold approach. This turns out {\em not}
to be the case in the ${\bf Z}_{12}$ model which we will discuss in detail in
a moment. However, some of the sectors in the ${\bf Z}_{12}$ model are similar
to those in the ${\bf Z}_6^*$ model, which was explicitly constructed 
in \cite{KS}, so let us discuss the latter in a bit more detail.

{}Thus, consider Type I on the generalized Voisin-Borcea orbifold
$(T^2\otimes T^2\otimes T^2)/{\bf Z}_6^*$, where the action of the 
generator $g$ of ${\bf Z}_6^*$ on the complex coordinates $z_1,z_2,z_3$ 
parametrizing the three two-tori is given by:
\begin{equation}
 gz_1=\alpha z_1~,~~~gz_{2,3}=-\alpha z_{2,3}~,
\end{equation} 
where $\alpha\equiv\exp(2\pi i/3)$. In this case we have 32 D9-branes and 32
D5-branes, the latter wrapping the first $T^2$. The non-perturbative open 
string sectors correspond to the orientifold group elements labeled by 
$\Omega g$ plus $\Omega g^5$ and $\Omega g^2$ plus $\Omega g^4$. Note that
the singularities in the ${\bf Z}_6^*$ twisted sectors are a subset of 
singularities in the ${\bf Z}_3^*$ twisted sectors. This implies that 
if the blow-ups in the ${\bf Z}_3^*$ model lead to the decoupling of 
non-perturbative open string sector states, the same should hold in the 
${\bf Z}_6^*$ model as well. The fact that non-perturbative twisted open string
sector states indeed decouple in the blown-up ${\bf Z}_3^*$ model was shown
in \cite{ZK} using Type-I heterotic duality. Thus, in the corresponding ${\bf 
Z}_6^*$ model we expect all non-perturbative states to decouple as well.

\subsection{The ${\bf Z}_{12}$ Model: A Non-perturbative Anomaly}

{}We are now ready to discuss the ${\bf Z}_{12}$ model. The generator $g$ 
of ${\bf
Z}_{12}$ has the following action on the three complex coordinates in the 
compact directions:
\begin{equation}
 gz_1=\alpha z_1~,~~~gz_2=i\alpha z_2~,~~~gz_3=-i\alpha z_3~. 
\end{equation} 
Note that ${\bf Z}_{12}\supset {\bf Z}_6^*$. The corresponding Calabi-Yau 
four-fold has the Hodge numbers $(h^{1,1},h^{2,1})=(29,5)$, so that the closed 
string spectrum contains 34 chiral supermultiplets. Here we note that the Hodge
numbers in the ${\bf Z}_6^*$ case are also $(h^{1,1},h^{2,1})=(29,5)$. That is,
in both the ${\bf Z}_6^*$ and ${\bf Z}_{12}$ models the compactification space
is the same, but the gauge bundles are different. 

{}In the open string sector we have 32 
D9-branes as well as 32 D5-branes (in the following we will focus on the 
brane configuration where all D5-branes are sitting on top of each other
at the orientifold 5-plane located at $z_2=z_3=0$). The perturbative open
string sector of this model can be obtained using the standard orientifold 
techniques (once the action of the orbifold group on the Chan-Paton factors
is specified - see below). 
The non-perturbative sectors corresponding to the orientifold group
elements labeled by $\Omega g$ plus $\Omega g^{11}$, $\Omega g^2$ plus $\Omega
g^{10}$, $\Omega g^4$ plus $\Omega g^8$, and 
$\Omega g^5$ plus $\Omega g^7$ do not give rise to massless states
(after the appropriate blow-ups) for the same reasons as in the ${\bf Z}_6^*$
case. However, the ${\bf Z}_4$ twisted open string sectors corresponding to the
orientifold group elements labeled by $\Omega g^3$ plus $\Omega g^9$ do give 
rise to non-perturbative massless states. These states can be obtained by
starting from the ${\bf Z}_4$ model in Table I and projecting onto the 
${\bf Z}_3^*$ invariant states. There is, however, a subtlety in this 
projection, so let us discuss the action of the orbifold group on the 
Chan-Paton factors in a bit more detail.

{}The point here is that to have a perturbative (from the orientifold
viewpoint) gauge bundle, we must make sure that all the twisted tadpole
cancellation conditions (derived in the perturbative orientifold approach)
cancel. In particular, the twisted tadpole cancellation condition for the
${\bf Z}_{12}$, ${\bf Z}_6^*$, ${\bf Z}_4$ and ${\bf Z}_2$ twisted Chan-Paton
matrices (for both D9- and D5-branes) read \cite{KS,ibanez}:
\begin{equation}
 {\mbox{Tr}}(\gamma_{g^k})=0~,~~~k=1,2,3,5,6,7,9,10,11~.
\end{equation}    
On the other hand, the ${\bf Z}_3^*$ twisted Chan-Paton matrices are fixed
by the corresponding tadpole cancellation conditions as in the ${\bf Z}_3^*$
model of \cite{Sagnotti} (see the next subsection for details). 
The only way to satisfy all of these twisted
tadpole cancellation conditions is to consider $32\times 32$ Chan-Paton 
matrices which (up to equivalent representations) are given by
\begin{eqnarray}
 &&\gamma_{g^4}={\mbox{diag}}(\alpha{\bf I}_{12},\alpha^{-1}{\bf I}_{12},
 {\bf I}_8)~,\\
 &&\gamma_{g^3}={\mbox{diag}}({\bf U}\otimes {\bf I}_3,{\bf U}\otimes 
 {\bf I}_3,{\bf U}\otimes{\bf I}_2)~,
\end{eqnarray} 
where ${\bf U}\equiv{\mbox{diag}}(\beta,-\beta,\beta^{-1},-\beta^{-1})$,
and $\beta\equiv\exp(\pi i/4)$. Note that the same would not be possible if
we considered $16\times 16$ matrices. That is, the tadpole cancellation 
conditions require that both 16 D-branes and their orientifold images 
participate in canceling the tadpoles at the same time, 
while separate cancellations
of the corresponding tadpoles for the 16 D-branes and their orientifold images
cannot be achieved. This, in particular, implies that (since the 16 D-branes
and their orientifold images are mapped to each other by the orientifold 
projection) the Chan-Paton matrix $\gamma_\Omega$ interchanges the twisted
Chan-Paton matrices with their conjugates: $\gamma_\Omega:\gamma_{g^k}
\rightarrow \gamma_{g^{-k}}$. Such an action is perfectly well defined
in the six dimensional $\Omega J^\prime$ orientifolds of \cite{GJ} as in those 
cases the orientifold projection $\Omega J^\prime$ maps the $g$-twisted sector
to its conjugate $g^{-1}$-twisted sector. 
However, as we have already pointed out in Introduction,
in four dimensions the orientifold projection is $\Omega$ which maps the 
$g$-twisted sector to itself (here we assume that the appropriate blow-ups
have been performed). It is then {\em a priori} far from being obvious
whether it is consistent to have $\gamma_\Omega:\gamma_{g^k}
\rightarrow \gamma_{g^{-k}}$ while $\Omega: g^k\rightarrow g^k$. 
In fact, we are going to argue that such an action leads to inconsistencies.
Here we should point out that such an inconsistency cannot be seen 
perturbatively in the orientifold picture if we just focus on the gauge 
(that is, the open string) sector of the theory. Indeed, the perturbative 
open string sector of this model by itself is perfectly sensible (see below).
However, if one couples open and closed strings, then one expects an 
inconsistency to show up - thus, the twisted closed string sector states
(which couple to the corresponding D-branes) feel the orientifold projection
in a way which is not compatible with the action of the orientifold projection
on the Chan-Paton factors. That is, the orientifold group actions on the closed
and open string degrees of freedom are not compatible. It is, however, 
difficult to see this open-closed coupling inconsistency directly as the
$\Omega$ projection in the closed string sector makes sense only after the
appropriate blow-ups have been performed, so we are no longer at the orbifold 
point in the corresponding 
Calabi-Yau moduli space, which makes world-sheet computations 
rather involved (and practically useless, at least for our purposes here).

{}There are, however, other ways to see this inconsistency. First, we can try
to understand it using Type I-heterotic duality. Here we should emphasize that
quantifying the following statements in this context is difficult as the 
heterotic dual of the ${\bf Z}_{12}$ model would be non-perturbative - we have
5-branes in this case. Nonetheless, the following somewhat hand-waving argument
might still be useful. Thus, on the heterotic side the embedding of the gauge
bundle (that is, of the orbifold action on the gauge degrees of freedom)
can be understood in terms of the appropriate ${\bf Z}_{12}$ valued
${\mbox{Spin(32)}}/{\bf Z}_2$ lattice shifts. (These shifts are ${\bf Z}_{12}$
valued w.r.t. the ${\mbox{Spin(32)}}/{\bf Z}_2$ lattice, but are ${\bf Z}_{24}$
valued w.r.t. the $SO(32)$ lattice.) The $\Omega$ projection on the heterotic
side can be thought of as pairing 32 real Majorana fermions (in the real
fermion representation of the ${\mbox{Spin(32)}}/{\bf Z}_2$ degrees of freedom)
into 16 complex world-sheet fermions. The ${\bf Z}_{12}$ valued 
${\mbox{Spin(32)}}/{\bf Z}_2$ lattice shifts in this language translate into
$U(1)^{16}$ phase rotations of the complex fermions. For these 
rotations to be consistent, we must make sure that they can be written in the
complex $16\times 16$ basis. If this cannot be achieved, that is, if
we can only do this in the $32\times 32$ bases, then the model is 
inconsistent\footnote{On the Type I side this would amount to an inconsistency
arising in the world-sheet theory of a probe D-string placed in this 
background. I would like to thank E. Gimon for a discussion in a related but
somewhat different context.}.
In particular, the scattering amplitudes would be inconsistent \cite{KLST}.  
In the following subsection we will relate this argument to the analogous 
statement in the dual F-theory picture, where the above inconsistency can be
seen in a geometric way. However, for now let us take the above gauge bundle 
and discuss the corresponding massless spectrum arising in the ${\bf Z}_{12}$ 
model. As we will see in a moment, the {\em non-perturbative} spectrum turns 
out to be anomalous.

{}Once the twisted Chan-Paton matrices are fixed as above, the perturbative
(from the orientifold viewpoint) states in the ${\bf Z}_{12}$ model
are straightforward to work out.
The resulting perturbative 
massless open string sector is given\footnote{Here we note that the 
perturbative open string spectrum of the ${\bf Z}_{12}$ model was discussed
in \cite{ibanez}. The untwisted open string spectrum given in Table IV, 
however, differs from that in \cite{ibanez}. In particular, note that the 
spectrum should be invariant under permuting the two $U(2)$ 
subgroups (accompanied by an appropriate interchange/conjugation
of the corresponding 
$U(3)$ subgroups) in, say, the 99 open string sector, which follows from the
action of the orbifold group on the Chan-Paton factors. 
This is the case for the spectrum in Table IV, but {\em not} for the spectrum
given in \cite{ibanez}.} in Table IV. Note that the non-Abelian 
gauge anomalies cancel in the untwisted open string 
sector. 

{}Next, let us discuss the non-perturbative open string sectors, that is, the 
${\bf Z}_4$ twisted open string sectors. These are straightforward to work out
starting from the ${\bf Z}_4$ twisted open string sectors in the ${\bf Z}_4$ 
model in Table I and projecting onto the ${\bf Z}_3^*$ invariant states. Note
that the resulting twisted 99 and 55 open string sectors are apparently T-dual 
for the same reason
as in the ${\bf Z}_4$ model in Table I - the ${\bf Z}_4$ fixed points are
invariant under the ${\bf Z}_2$ reflections involved in the T-duality 
transformation (recall that the other twisted sectors such as 
${\bf Z}_3^*$, ${\bf Z}_6^*$
and ${\bf Z}_{12}$ do not give rise to non-perturbative states once 
the appropriate blow-ups are performed). In fact, it is important here that
the ${\bf Z}_4$ sector fixed point located at the origin contains no twisted
$B$-field, whereas the other three fixed points have half-integer valued
$B$-field stuck inside of the corresponding ${\bf P}^1$'s (see section III
for details). In particular, this configuration possesses the required 
${\bf Z}_3$ symmetry. 

{}Next, we list the resulting non-perturbative states arising in the 
${\bf Z}_4$ twisted open string sectors (these states are ${\cal N}=1$
chiral supermultiplets - see Table IV for notations):
\begin{eqnarray}
 &&({\overline {\bf 3}},{\bf 1},{\bf 1},{\bf 1},{\bf 1},
 {\bf 1})_{99(T)/55(T)}~,\nonumber\\
 &&({\bf 1},{\bf 1},{\overline {\bf 3}},{\bf 1},{\bf 2},
 {\bf 1})_{99(T)/55(T)} ~,\nonumber\\
 &&({\bf 1},{\bf 1},{\overline {\bf 3}},{\bf 1},{\bf 1},
 {\bf 1})_{99(T)/55(T)}~,\nonumber\\
 &&({\overline {\bf 3}},{\bf 1},{\bf 1},{\bf 1},{\overline {\bf 2}},
 {\bf 1})_{99(T)/55(T)}~,\nonumber\\
 &&({\bf 1},{\overline {\bf 3}},{\bf 1},{\bf 1},{\bf 1},
 {\bf 1})_{99(T)/55(T)}~,\nonumber\\
 &&({\bf 1},{\bf 1},{\bf 1},{\overline {\bf 3}},{\bf 1},
 {\bf 2})_{99(T)/55(T)}~,\nonumber\\
 &&({\bf 1},{\bf 1},{\bf 1},{\overline {\bf 3}},{\bf 1},
 {\bf 1})_{99(T)/55(T)}~,\nonumber\\
 &&({\bf 1},{\overline {\bf 3}},{\bf 1},{\bf 1},{\bf 1},
 {\overline {\bf 2}})_{99(T)/55(T)}~,\nonumber
\end{eqnarray}
where the first four of these states come from the hypermultiplet
transforming in $({\bf 28},{\bf 1})$, and
the last four come from the hypermultiplet transforming in 
$({\bf 1},{\bf 28})$ upon ${\bf Z}_3^*$ projecting the 
$[U(8)\otimes U(8)]_{99/55}$ quantum numbers in the ${\bf Z}_4$ model in 
Table I. Note that the above twisted spectrum suffers from $SU(3)^4$ 
non-Abelian gauge anomalies. That is, non-perturbatively we see an 
inconsistency in this model. In the next subsection we will give additional
arguments clarifying the origins of this non-perturbative anomaly using the
map between Type IIB orientifolds and F-theory.  

\subsection{Map to F-theory}

{}In the previous section we gave the map between Type I on Voisin-Borcea
orbifolds and F-theory. In this subsection we would like to do the same in the
case of generalized Voisin-Borcea orbifolds with $N=3,4,6$. There is a 
subtlety in this map (as was pointed out in \cite{KST}) in the $N=3,6$ cases,
and here we will give make this map precise.

{}Naively, we can start from ${\cal Y}_3=(T^2\otimes {\cal W}_2)/{\bf Z}_N$,
T-dualize along the fibre $T^2$, and map the resulting model to F-theory
via \cite{sen} to arrive at the F-theory compactification on the following
Calabi-Yau four-fold:
\begin{equation}\label{four-fold}
 {\cal X}_4=({\widetilde T}^2\otimes T^2\otimes {\cal W}_2)/
 ({\bf Z}_2\otimes {\bf Z}_N)~,
\end{equation}   
where the action of the generator ${\widetilde S}$ of ${\bf Z}_2$ is given by 
${\widetilde S}{\widetilde z}_0=-{\widetilde z}_0$, ${\widetilde S}z_0=-z_0$, 
and it acts trivially on ${\cal W}_2$; the action of the generator $\eta$ of
${\bf Z}_N$ is given by $\eta {\widetilde z}_0={\widetilde z}_0$, $\eta z_0=
\omega z_0$, $\eta \Omega_2=\omega^{-1} \Omega_2$. Here ${\widetilde z}_0$ and
$z_0$ parametrize ${\widetilde T}^2$ respectively $T^2$, $\omega\equiv
\exp(2\pi i/N)$, and $\Omega_2$ is the holomorphic two-form on ${\cal W}_2$
(which is a K3 surface). Note, however, that the ${\bf Z}_N$ 
singularities in the ${\cal Y}_3$ fibre $T^2$ are invariant under 
${\widetilde S}$ only in the $N=2,4$ cases. Thus, there is no subtlety in the 
above map in these cases. In the $N=3,6$ cases, however, the ${\bf Z}_N$ 
singularities are not invariant under ${\widetilde S}$, so that we must be a 
bit more careful. In fact, the issue in the $N=6$ case is the same as in the
$N=3$ case, so we can focus on the latter. Moreover, instead of considering
the generalized Voisin-Borcea orbifolds ${\cal Y}_3$, we will discuss this 
point for a simpler example, namely, the ${\bf Z}_3^*$ orbifold. The 
discussion for the other cases is completely analogous. Thus, let us start from
Type I on ${\cal Y}_3=(T^2\otimes T^2\otimes T^2)/{\bf Z}_3^*$, where the 
generator $g$ of ${\bf Z}_3^*$ acts on the complex coordinates $z_1,z_2,z_3$
parametrizing the three two-tori as $gz_i=\alpha z_i$ ($i=1,2,3$, $\alpha\equiv
\exp(2\pi i/3)$). The Hodge numbers of this Calabi-Yau three-fold
are given by $(h^{1,1},h^{2,1})=(36,0)$. 

{}In this Type I compactification, which was studied in detail 
in \cite{Sagnotti,ZK,cvetic},
we have 32 D9-branes and no D5-branes. The twisted tadpole cancellation 
conditions read ${\mbox{Tr}}(\gamma_g)=-2$ (in the $16\times 16$ basis). This
implies that (up to equivalent representations) we have
\begin{equation}
 \gamma_g={\mbox{diag}}(\alpha{\bf I}_6,\alpha^{-1} {\bf I}_6,{\bf I}_4)~, 
\end{equation}
where $\alpha\equiv\exp(2\pi i/3)$. The gauge group of this model is 
$U(12)\otimes SO(8)$, and the (untwisted) 99 open string sector contains
the following chiral supermultiplets: $3\times ({\bf 66},{\bf 1})$ plus
$3\times ({\overline {\bf 12}},{\bf 8}_v)$.

{}Here we would like to ask what is the T-dual (where we T-dualize along both
directions in the first $T^2$) of this Type I background. That is, we would 
like to understand the dual Type IIB orientifold with D7-branes. In fact, to
T-dualize we must first turn on Wilson lines which break the original $SO(32)$
gauge group (before ${\bf Z}_3^*$ orbifolding) to $SO(8)^4$. Indeed, the 
corresponding T-dual of Type I on $T^2\otimes T^2\otimes T^2$ is the 
$\Omega {\widetilde J} (-1)^{F_L}$ orientifold of Type IIB on $T^2\otimes 
T^2\otimes T^2$, where ${\widetilde J} z_1=-z_1$, and ${\widetilde J} z_{2,3}=
z_{2,3}$. Note that $T^2/{\widetilde J}$ has four fixed points at which are 
located the corresponding orientifold 7-planes. We must place 8 D7-branes at 
each of these orientifold 7-planes to properly cancel tadpoles (this statement
is precise in the perturbative orientifold context). Thus, the gauge group
is $SO(8)^4$. In fact, let us see what are the locations of these fixed points.
In order for $T^2$'s to have ${\bf Z}_3^*$ symmetry (by which we are 
ultimately going to mod out), they must be hexagonal. Let us focus on the first
$T^2$ whose volume we will denote by $v_1$. Then the metric on $T^2$ is given 
by $g_{ab}=\sqrt{v_1/3} \, 
e_a\cdot e_b$, $a,b=1,2$, where $e_a$ are the vectors
spanning the $SU(3)$ lattice. That is, $e_1^2=e_2^2=-2e_1\cdot e_2=2$. The 
${\widetilde J}$ fixed points are located at $\xi_0=0$, $\xi_a=e_a/2$, 
$a=1,2,3$, where $e_3\equiv -(e_1+e_2)$.

{}The T-dual Type I configuration of the above Type IIB orientifold is given 
by the following. Start from Type I with 32 D9-branes on $T^2\otimes 
T^2\otimes T^2$. Turn on two Wilson lines - one on the $a$-cycle and the other
one on the $b$-cycle of the first $T^2$ - such that they break the $SO(32)$ 
gauge symmetry down to $SO(8)^4$. This is precisely the configuration we are
looking for. Note that this setup is symmetric under ${\bf Z}_3$ valued 
permutations of any three of the four $SO(8)$ subgroups accompanied by $2\pi/3$
rotations of the first $T^2$. Under these rotations the first Wilson line maps
to the second Wilson line (as the $a$-cycle maps to the $b$-cycle on $T^2$).
Therefore, we can mod out by the ${\bf Z}_3^*$ symmetry. The corresponding
gauge bundle is given by the following Chan-Paton matrix:
\begin{equation}
 \gamma_g={\mbox{diag}}({\bf P}_3\otimes {\bf I}_4,{\bf W})~,
\end{equation}      
where ${\bf P}_3$ is a $3\times 3$ matrix of cyclic permutations, and
${\bf W}\equiv{\mbox{diag}}(\alpha{\bf I}_2,\alpha^{-1}{\bf I}_2)$. Note that
we still have ${\mbox{Tr}}(\gamma_g)=-2$, so that the twisted tadpole 
cancellation conditions are satisfied. The resulting gauge group is
$SO(8)\otimes U(4)$, and the matter consists of chiral supermultiplets in
$3\times ({\bf 28},{\bf 1})$ plus $3\times ({\bf 1},{\bf 6})$. The $SO(8)$
factor arises as the diagonal subgroup of the original $SO(8)^3$ subgroup
on which the permutation matrix ${\bf P}_3$ is acting. The $U(4)$ factor is
the corresponding subgroup of the fourth $SO(8)$.

{}The above $SO(8)\otimes U(4)$ model actually is on the same moduli as the 
original $U(12)\otimes SO(8)$ model \cite{ZK}. The former point in the moduli
space is T-dual of the corresponding Type IIB orientifold with D7-branes, and
the latter can now be readily mapped to F-theory. The corresponding 
compactification space is given by the four-fold ${\cal X}_4=
({\widetilde T}^2\otimes T^2\otimes T^2\otimes T^2)/({\bf Z}_2\otimes 
{\bf Z}_3^*)$, which has $SU(4)$ holonomy, and is an elliptic fibration of 
${\widetilde T}^2$ over the base $(T^2\otimes T^2\otimes T^2)/({\bf Z}_2
\otimes {\bf Z}_3^*)$. 

{}We can generalize the above discussion to other examples of this type.
In particular, Type I on ${\cal Y}_3=(T^2\otimes {\cal W}_2)/{\bf Z}_N$
(where ${\cal W}_2$ is either a K3 surface or $T^4$) with the appropriately 
turned on Wilson lines is dual to F-theory on the elliptic 
four-fold (\ref{four-fold}). The Euler number of a four-fold is expressed in
terms of the corresponding Hodge numbers via:
\begin{equation}
 \chi=4+2h^{1,1}-4h^{2,1}+2h^{3,1}+h^{2,2}~.
\end{equation}
However, for elliptic four-folds we have 
\begin{equation}
 h^{2,2}=44+4h^{1,1}-2h^{2,1}+4h^{3,1}~,
\end{equation}  
so it suffices to give the Hodge numbers $(h^{1,1,},h^{2,1},h^{3,1})$ to
specify such a four-fold. In particular, we have 
\begin{equation}
 \chi/6=8+h^{1,1}-h^{2,1}+h^{3,1}~.
\end{equation}
For illustrative purposes, here we give 
the Hodge numbers of the four-folds corresponding to Type I
on the ${\bf Z}_3^*$, ${\bf Z}_6^*$, ${\bf Z}_{12}^*$, and ${\bf Z}_6^*\otimes
{\bf Z}_2$ orbifolds (other cases are straightforward to work out):
\begin{eqnarray}
 {\bf Z}_3^*:&~~~&(h^{1,1},h^{2,1},h^{3,1})=(32,21,9)~,~~~
 \chi/24=7~,\nonumber\\
 {\bf Z}_6^*,{\bf Z}_{12}:&~~~&(h^{1,1},h^{2,1},h^{3,1})=(32,6,14)~,~~~
 \chi/24=12~,\nonumber\\
 {\bf Z}_6^*\otimes {\bf Z}_2:&~~~&(h^{1,1},h^{2,1},h^{3,1})=(60,1,1)~,~~~
 \chi/24=17~.\nonumber
\end{eqnarray} 
Note that in the ${\bf Z}_6^*$ and ${\bf Z}_{12}$ cases we have the same 
four-fold. This is not surprising as the corresponding three-folds 
were also the same.

{}Finally, we would like to apply the above
map to F-theory to better understand the
non-perturbative inconsistency we have encountered in the ${\bf Z}_{12}$ case
with the particular gauge bundle discussed in the previous subsection. The
point here is that in the F-theory language the action of the orientifold 
projection $\Omega$ is geometrized - it is mapped to the action of 
${\widetilde J}$. Now, in the case of the four-fold compactifications of 
F-theory we must specify the action of ${\widetilde J}$ not only on the
four complex coordinates, but also on the gauge bundle. More
precisely, in, say, the ${\bf Z}_{12}$ case we have to specify the action of
the orbifold group ${\bf Z}_{12}$ on various gauge degrees of freedom, that is,
we have to specify the gauge bundle. In the orientifold language this is 
described by the Chan-Paton matrices, while in the heterotic language this is
done in terms of the ${\mbox{Spin(32)}}/{\bf Z}_2$ lattice shifts. In the 
F-theory context we must also specify the orbifold action of the corresponding
gauge degrees of freedom. Thus, in the ${\bf Z}_3^*$ example discussed above
the ${\bf Z}_3^*$ orbifold acts geometrically on the three $SO(8)$'s 
(corresponding to $D_4$ singularities in the fibre ${\widetilde T}^2$) by 
permuting them - this is precisely what happens to the three ${\widetilde J}$
fixed points $\xi_a=e_a/2$, $a=1,2,3$, under the action of ${\bf Z}_3^*$. 
However, the action of ${\bf Z}_3^*$ on the fourth $SO(8)$ 
(corresponding to the fixed point $\xi_0=0$) is no longer purely 
geometric - it breaks $SO(8)$ down to $U(4)$, which implies that there is 
non-trivial gauge bundle associated with the embedding of the ${\bf Z}_3^*$ 
orbifold action on the corresponding gauge degrees of freedom. The situation 
here is similar to what happens in the heterotic compactifications - we must 
embed some number of instantons which break the gauge group. Now, the action
of ${\widetilde J}$ on the ${\bf Z}_{12}$ twisted sectors {\em as well as} 
the corresponding gauge bundles must be one and the same representation
of the orbifold group. In the gauge bundle we discussed in the previous
subsection the orbifold group (in the four-fold context) is ${\bf Z}_2\otimes
{\bf Z}_{12}$, whereas its embedding in terms of the gauge bundle would be
the non-Abelian dihedral group $D_{12}$. These two actions are clearly 
incompatible, hence the non-perturbative inconsistency (which, in particular,
manifests itself via a non-perturbative anomaly) in the ${\bf Z}_{12}$ model.  

\section{Extensions}

In the previous sections we have discussed various non-perturbative 
orientifolds corresponding to Type I compactifications on K3 and Calabi-Yau 
three-folds. Our discussion so far has been confined to the cases with trivial
NS-NS antisymmetric $B$-field. It would be interesting to understand 
non-perturbative orientifolds with non-zero $B$-field, and one example of such
a compactification was recently discussed in \cite{NP}. We will not consider 
these models in detail in this paper as we are planning to discuss such
compactifications elsewhere \cite{new}. However, here we would like to point 
out an additional set of models arising in the {\em perturbative} K3 
orientifolds (which complement those discussed in \cite{bij}) as their 
generalizations to non-perturbative orientifolds might lead to interesting new 
models in six and four dimensions. 

{}The key point here is the following. Consider the $\Omega$ orientifold of
Type IIB on $T^4/{\bf Z}_2$ without the $B$-field. This is the model of 
\cite{PS,GP}. Note that the action of the orbifold group on the Chan-Paton
factors is given by the Chan-Paton matrix
\begin{equation}
 \gamma_g={\mbox{diag}}(i{\bf I}_8,-i{\bf I}_8)~,
\end{equation}
where $g$ is the generator of ${\bf Z}_2$. This corresponds to the gauge bundle
without vector structure (see section II for details). Here we can ask whether
we can take the Chan-Paton matrix to be given by
\begin{equation}
 \gamma_g={\mbox{diag}}({\bf I}_8,-{\bf I}_8)~,
\end{equation}
which would also satisfy the twisted tadpole cancellation conditions. In this 
case we would have the gauge bundle with vector structure. However, this choice
can be seen to be inconsistent. The point is that the gauge group in this
case is $[Sp(16)\otimes Sp(16)]_{99}\otimes [Sp(16)\otimes Sp(16)]_{55}$
(in our conventions $Sp(2N)$ has rank $N$), and the matter consists of
hypermultiplets in
\begin{eqnarray}
 && ({\bf 16},{\bf 16};{\bf 1},{\bf 1})_{99}~,~~~
 ({\bf 1},{\bf 1};{\bf 16},{\bf 16})_{55}~,\nonumber\\
 &&{1\over 2}({\bf 16},{\bf 1};{\bf 16},{\bf 1})_{59}~,~~~
 {1\over 2}({\bf 1},{\bf 16};{\bf 1},{\bf 16})_{59}~.\nonumber
\end{eqnarray} 
Note that in the 59 sectors we have half-hypermultiplets in {\em real}
representations, which is inconsistent, albeit the gravitational anomaly 
cancellation condition (\ref{anom}) is satisfied in this case (note that 
the number of extra tensor multiplets is zero).

{}Note, however, that, as was pointed out in \cite{bij}, if we turn on the
$B$-field of rank $b$ ($b=2,4$ in the case of K3 compactifications), the 59 
sector states come with multiplicity $2^{b/2}$ (while the rank of the gauge 
group is reduced by $2^{b/2}$ \cite{Bij,bij}). 
This implies that the 59 sector states no
longer need to come in half-hypermultiplets. Thus, consider the $b=2$
case. The Chan-Paton matrix is given by:
\begin{equation}\label{b=2}
 \gamma_g={\mbox{diag}}({\bf I}_4,-{\bf I}_4)~.
\end{equation}
The gauge group now is $[Sp(8)\otimes Sp(8)]_{99}\otimes 
[Sp(8)\otimes Sp(8)]_{55}$, and the matter consists of
hypermultiplets in
\begin{eqnarray}
 && ({\bf 8},{\bf 8};{\bf 1},{\bf 1})_{99}~,~~~
 ({\bf 1},{\bf 1};{\bf 8},{\bf 8})_{55}~,\nonumber\\
 &&({\bf 8},{\bf 1};{\bf 8},{\bf 1})_{59}~,~~~
 ({\bf 1},{\bf 8};{\bf 1},{\bf 8})_{59}~.\nonumber
\end{eqnarray}   
This spectrum is completely consistent, and, in particular, the gravitational
anomaly cancels (the number of extra tensor multiplets in this model is 4, 
which follows from the results of \cite{bij}). Note that this model was 
originally discussed in \cite{PS}.

{}In fact, we can generalize the above discussion to other perturbative 
K3 orientifolds, that is, the $\Omega J^\prime$ orientifolds of Type IIB on 
$T^4/{\bf Z}_N$, $N=2,3,4,6$. (Note that in the ${\bf Z}_2$ case the action of
$J^\prime$ is trivial.) Actually, in the ${\bf Z}_3$ case nothing changes
as we do not have a ${\bf Z}_2$ subgroup, but in other cases we do obtain new
models with non-zero $B$-field. Here we give the twisted Chan-Paton matrices
for these models, which we label $[N,b]$ ($g$ is the generator of 
${\bf Z}_N$):
\begin{eqnarray}
 &&[2,2]:~~~\gamma_g={\mbox{diag}}({\bf I}_4,-{\bf I}_4)~,\nonumber\\
 &&[2,4]:~~~\gamma_g={\mbox{diag}}({\bf I}_2,-{\bf I}_2)~,\nonumber\\
 &&[3,2]:~~~\gamma_g={\mbox{diag}}(\alpha{\bf I}_4,\alpha^{-1}
 {\bf I}_4)~,\nonumber\\
 &&[3,4]:~~~\gamma_g={\bf I}_4~,\nonumber\\
 &&[4,2]:~~~\gamma_g={\mbox{diag}}(i{\bf I}_2,-i{\bf I}_2,{\bf I}_2,
 -{\bf I}_2)~,\nonumber\\
 &&[4,4]:~~~\gamma_g={\mbox{diag}}(i,-i,1,-1)~,\nonumber\\
 &&[6,2]:~~~\gamma_g={\mbox{diag}}(\alpha {\bf I}_2,
 -\alpha {\bf I}_2,\alpha^{-1} {\bf I}_2,-\alpha^{-1}{\bf I}_2)~,\nonumber\\
 &&[6,4]:~~~\gamma_g={\mbox{diag}}({\bf I}_2,-{\bf I}_2)~,
\end{eqnarray}
where $\alpha\equiv\exp(2\pi i/3)$. These models, which were worked out in 
\cite{unpublished}, are summarized in Table V. Note that in all of these
models the massless spectra satisfy the gravitational anomaly cancellation 
condition (\ref{anom}). Here we note that the $[6,2]$ model in Table V is 
the same as the corresponding $[6,2]$ model without vector structure 
discussed in \cite{bij}. Also, the $[2,4]$ and $[6,4]$ models in Table V are
the same. Moreover, it is not difficult to show that each of the $[N,b]$ models
in Table V is on the same moduli as the corresponding $[N,b]$ model
without vector structure discussed in \cite{bij}.

\acknowledgments
 
{}This work was supported in part by the grant
NSF PHY-96-02074, 
and the DOE 1994 OJI award. I would also like to thank Albert and 
Ribena Yu for financial support.

\newpage
\begin{figure}[t]
\epsfxsize=16 cm
\epsfbox{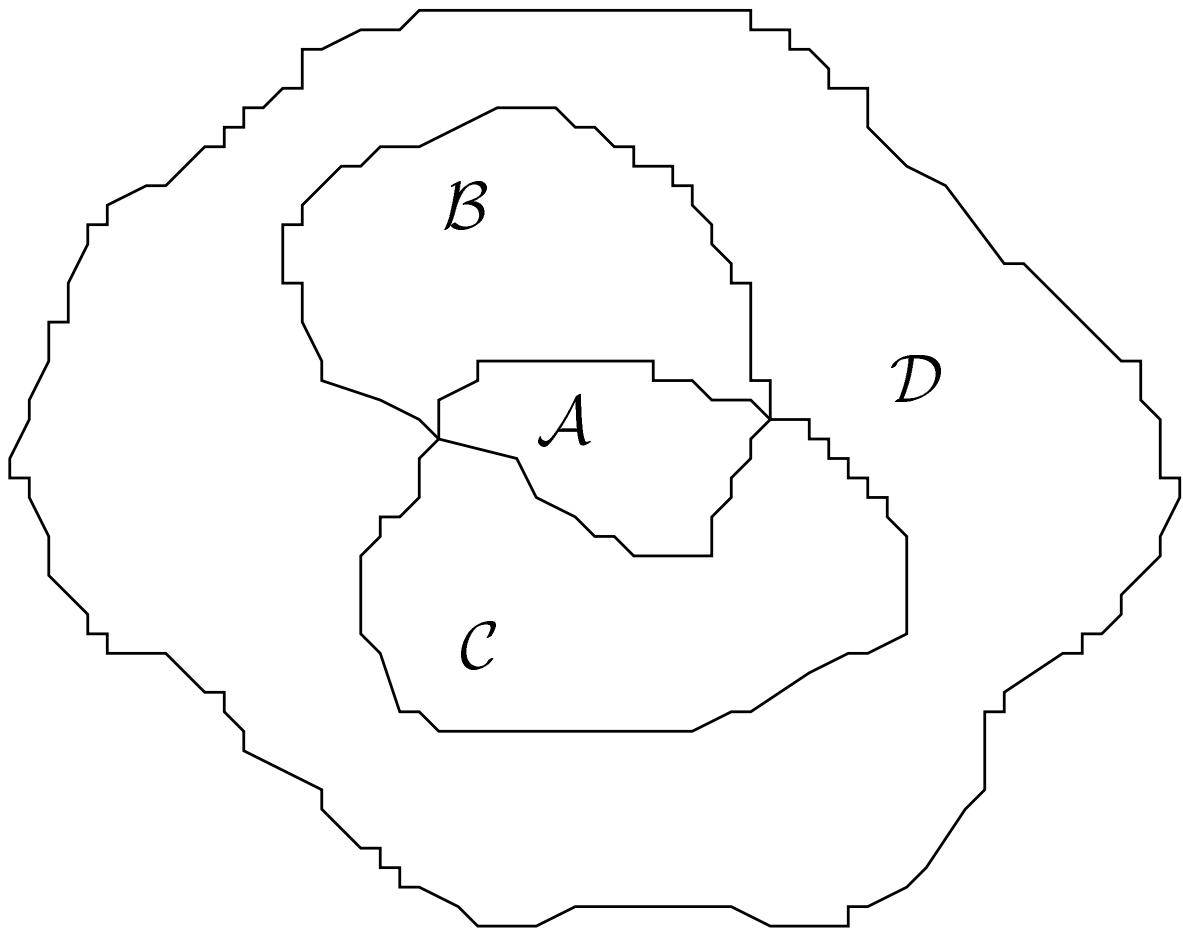}
\bigskip
\caption{A schematic picture of the space of four dimensional 
${\cal N}=1$ Type I and
heterotic vacua. The region ${\cal A}\cup{\cal B}$ corresponds to 
perturbative Type I vacua.
The region ${\cal A}\cup{\cal C}$ corresponds to perturbative heterotic 
vacua. The vacua
in the region ${\cal A}$ are perturbative from both the Type I and 
heterotic viewpoints. The region ${\cal D}$ contains both non-perturbative 
Type I and heterotic vacua.}
\end{figure}

\newpage
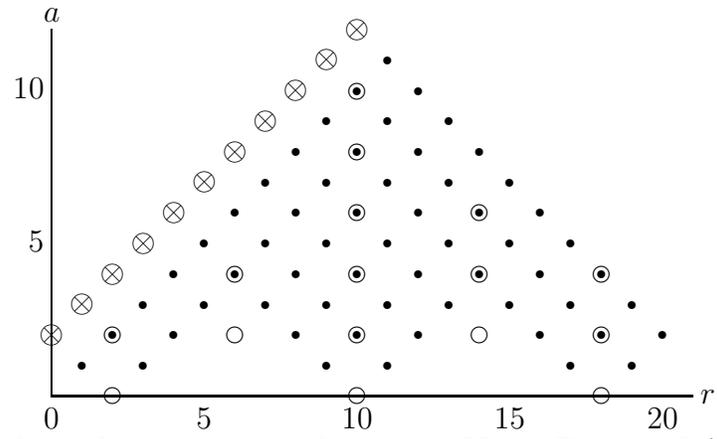
\begin{figure}
\setlength{\unitlength}{0.008in}%
$$\begin{picture}(445,266)(60,385)
\thinlines
\put(72,435){$\otimes$}
\put(92,455){$\otimes$}
\put(112,475){$\otimes$}
\put(132,495){$\otimes$}
\put(152,515){$\otimes$}
\put(172,535){$\otimes$}
\put(192,555){$\otimes$}
\put(212,575){$\otimes$}
\put(232,595){$\otimes$}
\put(252,615){$\otimes$}
\put(272,635){$\otimes$}

\put(100,420){\circle*{6}}

\put(140,420){\circle*{6}}
\put(260,420){\circle*{6}}
\put(300,420){\circle*{6}}
\put(420,420){\circle*{6}}
\put(460,420){\circle*{6}}
\put(120,440){\circle*{6}}
\put(160,440){\circle*{6}}
\put(240,440){\circle*{6}}
\put(280,440){\circle*{6}}
\put(320,440){\circle*{6}}
\put(400,440){\circle*{6}}
\put(440,440){\circle*{6}}
\put(140,460){\circle*{6}}
\put(180,460){\circle*{6}}
\put(220,460){\circle*{6}}
\put(260,460){\circle*{6}}
\put(300,460){\circle*{6}}
\put(340,460){\circle*{6}}
\put(380,460){\circle*{6}}
\put(420,460){\circle*{6}}
\put(160,480){\circle*{6}}
\put(200,480){\circle*{6}}
\put(240,480){\circle*{6}}
\put(280,480){\circle*{6}}
\put(360,480){\circle*{6}}
\put(400,480){\circle*{6}}
\put(180,500){\circle*{6}}
\put(220,500){\circle*{6}}
\put(260,500){\circle*{6}}
\put(300,500){\circle*{6}}
\put(340,500){\circle*{6}}
\put(380,500){\circle*{6}}
\put(200,520){\circle*{6}}
\put(240,520){\circle*{6}}
\put(280,520){\circle*{6}}
\put(320,520){\circle*{6}}
\put(360,520){\circle*{6}}
\put(220,540){\circle*{6}}
\put(260,540){\circle*{6}}
\put(300,540){\circle*{6}}
\put(340,540){\circle*{6}}
\put(240,560){\circle*{6}}
\put(280,560){\circle*{6}}
\put(320,560){\circle*{6}}
\put(260,580){\circle*{6}}
\put(300,580){\circle*{6}}
\put(280,600){\circle*{6}}
\put(320,480){\circle*{6}}
\put(120,400){\circle{10}}
\put(280,400){\circle{10}}
\put(440,400){\circle{10}}
\put(200,440){\circle{10}}
\put(200,480){\circle{10}}
\put(360,440){\circle{10}}
\put(360,480){\circle{10}}
\put(280,480){\circle{10}}
\put(280,520){\circle{10}}
\put(280,560){\circle{10}}
\put(280,600){\circle{10}}
\put(280,440){\circle{10}}
\put(120,440){\circle{10}}
\put(440,440){\circle{10}}
\put( 80,400){\line( 1, 0){420}}
\put( 80,400){\line( 0, 1){240}}
\put(300,620){\circle*{6}}
\put(320,600){\circle*{6}}
\put(340,580){\circle*{6}}
\put(380,540){\circle*{6}}
\put(400,520){\circle*{6}}
\put(420,500){\circle*{6}}
\put(440,480){\circle*{6}}
\put(440,480){\circle{10}}
\put(460,460){\circle*{6}}
\put(480,440){\circle*{6}}
\put(360,520){\circle{10}}
\put(360,560){\circle*{6}}
\put( 75,379){\makebox(0,0)[lb]{\raisebox{0pt}[0pt][0pt]{0}}}
\put(175,379){\makebox(0,0)[lb]{\raisebox{0pt}[0pt][0pt]{5}}}
\put(270,379){\makebox(0,0)[lb]{\raisebox{0pt}[0pt][0pt]{10}}}
\put(370,379){\makebox(0,0)[lb]{\raisebox{0pt}[0pt][0pt]{15}}}
\put(470,379){\makebox(0,0)[lb]{\raisebox{0pt}[0pt][0pt]{20}}}
\put( 65,495){\makebox(0,0)[lb]{\raisebox{0pt}[0pt][0pt]{5}}}
\put( 55,595){\makebox(0,0)[lb]{\raisebox{0pt}[0pt][0pt]{10}}}
\put(505,395){\makebox(0,0)[lb]{\raisebox{0pt}[0pt][0pt]{$r$}}}
\put( 75,645){\makebox(0,0)[lb]{\raisebox{0pt}[0pt][0pt]{$a$}}}
\end{picture}$$
	\caption{Open circles and dots represent the original
                         Voisin--Borcea orbifolds.
                         The line of $\otimes$'s corresponds to the extension
                         discussed in section IV.}
	\label{figVB}
\end{figure}

\begin{table}[t]
\begin{tabular}{|c|c|l|c|c|}
Model & Gauge Group & \phantom{Hy} Charged  & Neutral 
& Extra Tensor  \\
     &             &Hypermultiplets & Hypermultiplets
&Multiplets \\
\hline
${\bf Z}_2$ & $U(16)_{99} \otimes U(16)_{55}$ & 
 $2 \times ({\bf 120};{\bf 1})_U$ & $20$
& $0$ \\
               &                       & $2 \times({\bf 1};{\bf 120})_U$ & & \\
               &                       & $({\bf 16};{\bf 16})_U$ & & \\
\hline
${\bf Z}_3$ & $[U(8) \otimes SO(16)]_{99}$ & $({\bf 28},{\bf 1})_U$ & $20$
& $0$ \\
& & $({\bf 8},{\bf 16})_U$ & & \\
& & $9\times ({\bf 28},{\bf 1})_T$ & & \\
\hline
${\bf Z}_4$ & $[U(8) \otimes U(8)]_{99}\otimes$ 
& $({\bf 28},{\bf 1};{\bf 1},{\bf 1})_U$ & $20$ & $0$ \\
& $[U(8) \otimes U(8)]_{55}$&
$({\bf 1},{\bf 28};{\bf 1},{\bf 1})_U$  & & \\
& & $({\bf 8},{\bf 8};{\bf 1},{\bf 1})_U$ & & \\
& & same as above with $99\leftrightarrow 55$ & & \\
& & $({\bf 8},{\bf 1};{\bf 8},{\bf 1})_U$ & & \\
& & $({\bf 1},{\bf 8};{\bf 1},{\bf 8})_U$ & & \\
& & $({\bf 28},{\bf 1};{\bf 1},{\bf 1})_T$ & & \\
& & $({\bf 1},{\bf 28};{\bf 1},{\bf 1})_T$ & & \\
& & $({\bf 1},{\bf 1};{\bf 28},{\bf 1})_T$ & & \\
& & $({\bf 1},{\bf 1};{\bf 1},{\bf 28})_T$ & & \\
\hline
${\bf Z}_6$ & $[U(4) \otimes U(4) \otimes U(8)]_{99}\otimes$ 
& $({\bf 6},{\bf 1},{\bf 1};{\bf 1},{\bf 1},{\bf 1})_U$ & $20$ & $0$ \\
& $[U(4) \otimes U(4) \otimes U(8)]_{55}$&
$({\bf 1},{\bf 6},{\bf 1};{\bf 1},{\bf 1},{\bf 1})_U$  & & \\
& & $({\bf 4},{\bf 1},{\bf 8};{\bf 1},{\bf 1},{\bf 1})_U$ & & \\
& & $({\bf 1},{\bf 4},{\bf 8};{\bf 1},{\bf 1},{\bf 1})_U$ & & \\
& & same as above with $99\leftrightarrow 55$ & &\\
& & $({\bf 4},{\bf 1},{\bf 1};{\bf 4},{\bf 1},{\bf 1})_U$ & & \\
& & $({\bf 1},{\bf 4},{\bf 1};{\bf 1},{\bf 4},{\bf 1})_U$ & & \\
& & $({\bf 1},{\bf 1},{\bf 8};{\bf 1},{\bf 1},{\bf 8})_U$ & & \\
& & $5\times ({\bf 6},{\bf 1},{\bf 1};{\bf 1},{\bf 1},{\bf 1})_T$ & & \\
& & $5\times ({\bf 1},{\bf 6},{\bf 1};{\bf 1},{\bf 1},{\bf 1})_T$ & & \\
& & $4\times ({\bf 4},{\bf 4},{\bf 1};{\bf 1},{\bf 1},{\bf 1})_T$ & & \\
& & $({\bf 1},{\bf 1},{\bf 1};{\bf 6},{\bf 1},{\bf 1})_T$ & & \\
& & $({\bf 1},{\bf 1},{\bf 1};{\bf 1},{\bf 6},{\bf 1})_T$ & & \\
& & $({\bf 4},{\bf 1},{\bf 1};{\bf 4},{\bf 1},{\bf 1})_T$ & & \\
& & $({\bf 1},{\bf 4},{\bf 1};{\bf 1},{\bf 4},{\bf 1})_T$ & & \\
\hline
\end{tabular}
\caption{The massless spectra of the six dimensional ${\cal N}=1$ 
supersymmetric Type IIB orientifolds
on $T^4/{\bf Z}_N$ for $N=2,3,4,6$.
The semi-colon in the column ``Charged Hypermultiplets'' separates $99$ and 
$55$ representations. The subscript ``$U$'' indicates that the
corresponding (``untwisted'') state is perturbative from the orientifold
viewpoint. The subscript ``$T$'' indicates that the
corresponding (``twisted'') state is non-perturbative from the orientifold
viewpoint. The $U(1)$ charges are not shown, and by ``neutral''
hypermultiplets we mean that the corresponding states are not charged
under the non-Abelian subgroups.}
\end{table}

\begin{table}[t]
\begin{tabular}{|c|c|l|c|}
Model & Gauge Group & \phantom{Hy} Charged  & Neutral 
 \\
     &             &Chiral Multiplets & Chiral Multiplets
\\
\hline
${\bf Z}_6^\prime$ & $[U(4) \otimes U(4) \otimes U(8)]_{99}\otimes$ 
& $({\bf 1},{\bf 1},{\bf 28};{\bf 1},{\bf 1},{\bf 1})_U$ & $46$ \\
& $[U(4) \otimes U(4) \otimes U(8)]_{55}$&
$({\bf 1},{\bf 1},{\overline {\bf 28}};
 {\bf 1},{\bf 1},{\bf 1})_U$  & \\
& & $({\bf 4},{\bf 4},{\bf 1};{\bf 1},{\bf 1},{\bf 1})_U$ &   \\
& & $({\overline {\bf 4}},{\overline {\bf 4}},{\bf 1};
 {\bf 1},{\bf 1},{\bf 1})_U$ &  \\
& & $({\bf 4},{\bf 1},{\bf 8};
 {\bf 1},{\bf 1},{\bf 1})_U$ &  \\
& & $({\bf 1},{\overline {\bf 4}},{\overline {\bf 8}};
 {\bf 1},{\bf 1},{\bf 1})_U$ &  \\
& & $({\overline {\bf 6}},{\bf 1},{\bf 1};
 {\bf 1},{\bf 1},{\bf 1})_U$ &  \\
& & $({\bf 1},{\bf 6},{\bf 1};
 {\bf 1},{\bf 1},{\bf 1})_U$ &  \\
& & $({\overline {\bf 4}},{\bf 1},{\bf 8};
 {\bf 1},{\bf 1},{\bf 1})_U$ &  \\
& & $({\bf 1},{\bf 4},{\overline {\bf 8}};
 {\bf 1},{\bf 1},{\bf 1})_U$ &  \\
& & $({\bf 4},{\overline {\bf 4}},{\bf 1};
 {\bf 1},{\bf 1},{\bf 1})_U$ &  \\
& & Same as above with $99\leftrightarrow 55$ & \\
& & $({\bf 4},{\bf 1},{\bf 1};{\bf 4},{\bf 1},{\bf 1})_U$ & \\
& & $({\bf 1},{\bf 4},{\bf 1};{\bf 1},{\bf 1},{\bf 8})_U$ &  \\
& & $({\bf 1},{\bf 1},{\bf 8};{\bf 1},{\bf 4},{\bf 1})_U$ & \\
& & $({\overline {\bf 4}},{\bf 1},{\bf 1};{\bf 1},{\bf 1},
 {\overline {\bf 8}})_U$ & \\
& & $({\bf 1},{\overline {\bf 4}},{\bf 1};{\bf 1},{\overline {\bf 4}},
 {\bf 1})_U$ &  \\
& & $({\bf 1},{\bf 1},{\overline {\bf 8}};{\overline {\bf 4}},{\bf 1},
 {\bf 1})_U$ & \\
& & $6\times ({\bf 4},{\overline {\bf 4}},{\bf 1};
 {\bf 1},{\bf 1},{\bf 1})_T$ &  \\
& & $3\times ({\overline {\bf 4}},{\bf 4},{\bf 1};
 {\bf 1},{\bf 1},{\bf 1})_T$ &  \\
& & $6\times ({\overline {\bf 6}},{\bf 1},{\bf 1};
 {\bf 1},{\bf 1},{\bf 1})_T$ &  \\
& & $3\times ({\bf 6},{\bf 1},{\bf 1};
 {\bf 1},{\bf 1},{\bf 1})_T$ &  \\
& & $6\times ({\bf 1},{\bf 6},{\bf 1};
 {\bf 1},{\bf 1},{\bf 1})_T$ &  \\
& & $3\times ({\bf 1},{\overline {\bf 6}},{\bf 1};
 {\bf 1},{\bf 1},{\bf 1})_T$ &  \\
& & $3\times ({\bf 1},{\bf 1},{\bf 1};
 {\bf 4},{\overline {\bf 4}},{\bf 1})_T$ &  \\
& & $3\times ({\bf 1},{\bf 1},{\bf 1};
 {\overline {\bf 6}},{\bf 1},{\bf 1})_T$ &  \\
& & $3\times ({\bf 1},{\bf 1},{\bf 1};{\bf 1},{\bf 6},{\bf 1})_T$ &  \\

& & $3\times ({\overline {\bf 4}},{\bf 1},{\bf 1};{\overline {\bf 4}},
 {\bf 1},{\bf 1})_T$ & \\
& & $3\times ({\bf 1},{\bf 4},{\bf 1};{\bf 1},{\bf 4},{\bf 1})_T$ &  \\
\hline
\end{tabular}
\caption{The massless spectrum of the four dimensional ${\cal N}=1$ 
supersymmetric Type IIB orientifold
on $T^6/{\bf Z}_6^\prime$.
The semi-colon in the column ``Charged Chiral Multiplets'' separates $99$ and 
$55$ representations. The subscript ``$U$'' indicates that the
corresponding (``untwisted'') state is perturbative from the orientifold
viewpoint. The subscript ``$T$'' indicates that the
corresponding (``twisted'') state is non-perturbative from the orientifold
viewpoint. The $U(1)$ charges are not shown, and by ``neutral''
chiral multiplets we mean that the corresponding states are not charged
under the non-Abelian subgroups.}
\end{table}

\begin{table}[t]
\begin{tabular}{|c|c|l|c|}
Model & Gauge Group & \phantom{Hy} Charged  & Neutral 
 \\
     &             &Chiral Multiplets & Chiral Multiplets
\\
\hline
${\bf Z}_2\otimes {\bf Z}_6$ & $[U(4) \otimes Sp(8)]_{99}\otimes$ 
& $({\bf 16},{\bf 1})_{99(U)/5_i5_i(U)}$ & $54$ \\
& $\bigotimes_{i=1}^3 [U(4) \otimes Sp(8)]_{5_i 5_i}$&
$({\bf 1},{\bf 28})_{99(U)/5_i5_i(U)}$  & \\
& & $({\bf 4}+{\overline {\bf 4}},{\bf 8})_{99(U)/5_i5_i(U)}$ &    \\
& & $({\bf 6}+{\overline {\bf 6}},{\bf 1})_{99(U)/5_i5_i(U)}$ &   \\
& & $({\bf 4},{\bf 1};{\overline {\bf 4}},{\bf 1})_{95_1(U)/5_2 5_3(U)}$ &  \\
& & $({\overline {\bf 4}},{\bf 1};{\bf 4},{\bf 1})_{95_1(U)/5_2 5_3(U)}$ &  \\
& & $({\bf 1},{\bf 8};{\bf 1},{\bf 8})_{95_1(U)/5_2 5_3(U)}$ &  \\
& & $({\bf 4},{\bf 1};{\bf 4},{\bf 1})_{95_{2}(U)/5_1 5_{3}(U)}$ &  \\
& & $({\overline {\bf 4}},{\bf 1};{\bf 1},{\bf 8})_{95_{2}(U)/5_1 5_{3}
 (U)}$ &  \\
& & $({\bf 1},{\bf 8};{\overline {\bf 4}},{\bf 1})_{95_{2}(U)/5_1 5_{3}
 (U)}$ &  \\
& & $({\overline {\bf 4}},{\bf 1};{\overline {\bf 4}},
 {\bf 1})_{95_{3}(U)/5_1 5_{2}(U)}$ &  \\
& & $({{\bf 4}},{\bf 1};{\bf 1},{\bf 8})_{95_{3}(U)/5_1 5_{2}
 (U)}$ &  \\
& & $({\bf 1},{\bf 8};{{\bf 4}},{\bf 1})_{95_{3}(U)/5_1 5_{2}
 (U)}$ &  \\
& & $2\times ({\bf 10}+{\overline {\bf 10}},{\bf 1})_{99(T)}$ &    \\
& & $7\times ({\bf 6}+{\overline {\bf 6}},{\bf 1})_{99(T)}$ &    \\
& & $({\bf 6}+{\overline {\bf 6}},{\bf 1})_{5_1 5_1(T)}$ &   \\
& & $({\bf 10}/{\overline {\bf 10}},{\bf 1})_{5_2 5_2(T)/5_3 5_3(T)}$ &    \\
& & $2\times ({\bf 6}/{\overline {\bf 6}},
 {\bf 1})_{5_2 5_2(T)/5_3 5_3(T)}$ &    \\
& & $3\times ({\overline {\bf 6}}/{\bf 6},
 {\bf 1})_{5_2 5_2(T)/5_3 5_3(T)}$ &    \\
& & $({\bf 4},{\bf 1};{\overline {\bf 4}},{\bf 1})_{95_1(T)/5_2 5_3(T)}$ &  \\
& & $({\overline {\bf 4}},{\bf 1};{\bf 4},{\bf 1})_{95_1(T)/5_2 5_3(T)}$ &  \\
& & $3\times ({\overline {\bf 4}},{\bf 1};{\overline {\bf 4}},
{\bf 1})_{95_{2}(T)}$ 
&  \\
& & $3\times ({{\bf 4}},{\bf 1};{{\bf 4}},
{\bf 1})_{95_{3}(T)}$ 
&  \\
& & $({{\bf 4}},{\bf 1};{{\bf 4}},
 {\bf 1})_{5_1 5_{2}(T)}$ 
&  \\
& & $({\overline {\bf 4}},{\bf 1};{\overline {\bf 4}},
 {\bf 1})_{5_1 5_{3}(T)}$ 
&  \\
\hline
\end{tabular}
\caption{The massless spectrum of the four dimensional ${\cal N}=1$ 
supersymmetric Type IIB orientifold
on $T^6/({\bf Z}_2\otimes {\bf Z}_6)$.
The semi-colon in the column ``Charged Chiral Multiplets'' separates $99$ and 
the corresponding $5_i5_i$ 
representations. The subscript ``$U$'' indicates that the
corresponding (``untwisted'') state is perturbative from the orientifold
viewpoint. The subscript ``$T$'' indicates that the
corresponding (``twisted'') state is non-perturbative from the orientifold
viewpoint. The $U(1)$ charges are not shown, and by ``neutral''
chiral multiplets we mean that the corresponding states are not charged
under the non-Abelian subgroups. Note that ${\bf 16}$ and ${\bf 28}$ are
{\em reducible} representations of $SU(4)$ and $Sp(8)$, respectively
(in our conventions $Sp(2N)$ has rank $N$).}
\end{table}

\begin{table}[t]
\begin{tabular}{|c|c|l|c|}
Model & Gauge Group & \phantom{Hy} Charged  & Neutral 
 \\
     &             &Chiral Multiplets & Chiral Multiplets
\\
\hline
${\bf Z}_{12}$ & $[U(3)^4 \otimes U(2)^2]_{99}\otimes$ 
& $({\bf 3},{\bf 1},{\bf 3},{\bf 1},{\bf 1},{\bf 1})_{99(U)/55(U)}$ & $34$ \\
& $[U(3)^4 \otimes U(2)^2]_{55}$&
$({\bf 1},{\bf 3},{\bf 1},{\bf 3},{\bf 1},{\bf 1})_{99(U)/55(U)}$  & \\
& & $({\overline {\bf 3}},{\bf 1},{\bf 1},{\bf 1},{\bf 2},
 {\bf 1})_{99(U)/55(U)}$ & \\
& & $({\bf 1},{\overline {\bf 3}},{\bf 1},
 {\bf 1},{\bf 1},{\bf 2})_{99(U)/55(U)}$ & \\
& & $({\bf 1},{\bf 1},{\overline {\bf 3}},
 {\bf 1},{\overline {\bf 2}},{\bf 1})_{99(U)/55(U)}$ &  \\
& & $({\bf 1},{\bf 1},{\bf 1},{\overline {\bf 3}},
 {\bf 1},{\overline {\bf 2}})_{99(U)/55(U)}$ &  \\
& & & \\
& & $({\overline {\bf 3}},{\bf 1},{\bf 1},{\bf 1},{\bf 1},
 {\bf 1})_{99(U)/55(U)}$ & \\
& &
$({\bf 1},{\overline {\bf 3}},{\bf 1},{\bf 1},{\bf 1},
 {\bf 1})_{99(U)/55(U)}$  & \\
& & $({\bf 1},{\bf 1},{\bf 3},{\bf 3},{\bf 1},
 {\bf 1})_{99(U)/55(U)}$ & \\
& & $({\overline {\bf 3}},{\bf 1},{\bf 1},
 {\bf 1},{\bf 1},{\overline {\bf 2}})_{99(U)/55(U)}$ & \\
& & $({\bf 1},{\overline {\bf 3}},{\bf 1},
 {\bf 1},{\overline {\bf 2}},{\bf 1})_{99(U)/55(U)}$ &  \\
& & $({\bf 1},{\bf 1},{\overline {\bf 3}},{\bf 1},
 {\bf 2},{\bf 1})_{99(U)/55(U)}$ &  \\
& & $({\bf 1},{\bf 1},{\bf 1},{\overline {\bf 3}},
 {\bf 1},{\bf 2})_{99(U)/55(U)}$ &  \\
& & &\\
& & $({\bf 1},{\bf 1},{\overline {\bf 3}},{\bf 1},{\bf 1},
 {\bf 1})_{99(U)/55(U)}$ & \\
& &
$({\bf 1},{\bf 1},{\bf 1},{\overline {\bf 3}},{\bf 1},
 {\bf 1})_{99(U)/55(U)}$  & \\
& & $({\bf 3},{\bf 3},{\bf 1},{\bf 1},{\bf 1},
 {\bf 1})_{99(U)/55(U)}$ & \\
& & $({\overline {\bf 3}},{\bf 1},{\bf 1},
 {\bf 1},{\overline {\bf 2}},{\bf 1})_{99(U)/55(U)}$ & \\
& & $({\bf 1},{\overline {\bf 3}},{\bf 1},
 {\bf 1},{\bf 1},{\overline {\bf 2}})_{99(U)/55(U)}$ &  \\
& & $({\bf 1},{\bf 1},{\overline {\bf 3}},{\bf 1},
 {\bf 1},{\bf 2})_{99(U)/55(U)}$ &  \\
& & $({\bf 1},{\bf 1},{\bf 1},{\overline {\bf 3}},
 {\bf 2},{\bf 1})_{99(U)/55(U)}$ &  \\
& & &\\
& & $({\bf 3},{\bf 1},{\bf 1},{\bf 1},
 {\bf 1},{\bf 1};{\bf 1},{\bf 1},{\bf 3},{\bf 1},
 {\bf 1},{\bf 1})_{59(U)}$ &  \\
& & $({\bf 1},{\bf 3},{\bf 1},{\bf 1},
 {\bf 1},{\bf 1};{\bf 1},{\bf 1},{\bf 1},{\bf 3},
 {\bf 1},{\bf 1})_{59(U)}$ &  \\
& & $({\bf 1},{\bf 1},{\bf 3},{\bf 1},
 {\bf 1},{\bf 1};{\bf 3},{\bf 1},{\bf 1},{\bf 1},
 {\bf 1},{\bf 1})_{59(U)}$ &  \\
& & $({\bf 1},{\bf 1},{\bf 1},{\bf 3},
 {\bf 1},{\bf 1};{\bf 1},{\bf 3},{\bf 1},{\bf 1},
 {\bf 1},{\bf 1})_{59(U)}$ &  \\
& & $({\overline {\bf 3}},{\bf 1},{\bf 1},{\bf 1},
 {\bf 1},{\bf 1};{\bf 1},{\bf 1},{\bf 1},{\bf 1},
 {\bf 2},{\bf 1})_{59(U)}$ &  \\
& & $({\bf 1},{\overline {\bf 3}},{\bf 1},{\bf 1},
 {\bf 1},{\bf 1};{\bf 1},{\bf 1},{\bf 1},{\bf 1},
 {\bf 1},{\bf 2})_{59(U)}$ &  \\
& & $({\bf 1},{\bf 1},{\overline {\bf 3}},{\bf 1},
 {\bf 1},{\bf 1};{\bf 1},{\bf 1},{\bf 1},{\bf 1},
 {\overline {\bf 2}},{\bf 1})_{59(U)}$ &  \\
& & $({\bf 1},{\bf 1},{\bf 1},{\overline {\bf 3}},
 {\bf 1},{\bf 1};{\bf 1},{\bf 1},{\bf 1},{\bf 1},
 {\bf 1},{\overline {\bf 2}})_{59(U)}$ &  \\
& & $({\bf 1},{\bf 1},{\bf 1},{\bf 1},
 {\bf 2},{\bf 1};{\overline {\bf 3}},{\bf 1},{\bf 1},{\bf 1},
 {\bf 1},{\bf 1})_{59(U)}$ &  \\
& & $({\bf 1},{\bf 1},{\bf 1},{\bf 1},
 {\bf 1},{\bf 2};{\bf 1},{\overline {\bf 3}},{\bf 1},{\bf 1},
 {\bf 1},{\bf 1})_{59(U)}$ &  \\
& & $({\bf 1},{\bf 1},{\bf 1},{\bf 1},
 {\overline {\bf 2}},{\bf 1};{\bf 1},{\bf 1},{\overline {\bf 3}},{\bf 1},
 {\bf 1},{\bf 1})_{59(U)}$ &  \\
& & $({\bf 1},{\bf 1},{\bf 1},{\bf 1},
 {\bf 1},{\overline {\bf 2}};{\bf 1},{\bf 1},{\bf 1},{\overline {\bf 3}},
 {\bf 1},{\bf 1})_{59(U)}$ &  \\
\hline
\end{tabular}
\caption{The {\em perturbative} 
massless spectrum of the four dimensional ${\cal N}=1$ 
supersymmetric Type IIB orientifold
on $T^6/{\bf Z}_{12}$.
The semi-colon in the column ``Charged Chiral Multiplets'' separates $99$ and 
$55$ representations. The subscript ``$U$'' indicates that the
corresponding (``untwisted'') state is perturbative from the orientifold
viewpoint. The $U(1)$ charges are not shown, and by ``neutral''
chiral multiplets we mean that the corresponding states are not charged
under the non-Abelian subgroups.}
\end{table}

\begin{table}[t]
\begin{tabular}{|c|c|c|l|c|c|}
Model & $b$ & Gauge Group & \phantom{Hy} Charged  & Neutral 
& Extra Tensor  \\
       &   &             &Hypermultiplets & Hypermultiplets
&Multiplets \\
\hline
${\bf Z}_2$ & 2 & $[Sp(8) \otimes Sp(8)]_{99}$ & $({\bf 8},{\bf 8};
 {\bf 1},{\bf 1})$ & $16$ & $4$ \\
& &$[Sp(8) \otimes Sp(8)]_{55}$ & $({\bf 1},{\bf 1};{\bf 8},{\bf 8})$ & & \\
& & & $({\bf 8},{\bf 1};{\bf 8},{\bf 1})$ & & \\
& & & $({\bf 1},{\bf 8};{\bf 1},{\bf 8})$ & & \\
\hline
& 4 & $[Sp(4) \otimes Sp(4)]_{99}$ & $({\bf 4},{\bf 4};
 {\bf 1},{\bf 1})$ & $14$ & $6$ \\
& &$[Sp(4) \otimes Sp(4)]_{55}$ & $({\bf 1},{\bf 1};{\bf 4},{\bf 4})$ & & \\
& &  & $2\times ({\bf 4},{\bf 1};{\bf 4},{\bf 1})$ & & \\
& & & $2\times ({\bf 1},{\bf 4};{\bf 1},{\bf 4})$ & & \\
\hline
${\bf Z}_4$ & 2 & $[U(4) \otimes Sp(4) \otimes Sp(4)]_{99}\otimes$ 
&$({\bf 4},{\bf 4},{\bf 1};{\bf 1},{\bf 1},{\bf 1})$ & $14$ & $6$ \\
& & $[U(4) \otimes Sp(4) \otimes Sp(4)]_{55}$&
$({\bf 4},{\bf 1},{\bf 4};{\bf 1},{\bf 1},{\bf 1})$  & & \\
& & & $({\bf 1},{\bf 1},{\bf 1};{\bf 4},{\bf 4},{\bf 1})$ & & \\
& & & $({\bf 1},{\bf 1},{\bf 1};{\bf 4},{\bf 1},{\bf 4})$ & & \\
& & & $({\bf 4},{\bf 1},{\bf 1};{\bf 1},{\bf 4},{\bf 1})$ & & \\
& & & $({\bf 4},{\bf 1},{\bf 1};{\bf 1},{\bf 1},{\bf 4})$ & & \\
& & & $({\bf 1},{\bf 4},{\bf 1};{\bf 4},{\bf 1},{\bf 1})$ & & \\
& & & $({\bf 1},{\bf 1},{\bf 4};{\bf 4},{\bf 1},{\bf 1})$ & & \\
\hline
& 4 & $[U(2) \otimes Sp(2) \otimes Sp(2)]_{99}\otimes$ 
&$({\bf 2},{\bf 2},{\bf 1};{\bf 1},{\bf 1},{\bf 1})$ & $13$ & $7$ \\
& & $[U(2) \otimes Sp(2) \otimes Sp(2)]_{55}$&
$({\bf 2},{\bf 1},{\bf 2};{\bf 1},{\bf 1},{\bf 1})$  & & \\
& & & $({\bf 1},{\bf 1},{\bf 1};{\bf 2},{\bf 2},{\bf 1})$ & & \\
& & & $({\bf 1},{\bf 1},{\bf 1};{\bf 2},{\bf 1},{\bf 2})$ & & \\
& & & $2\times ({\bf 2},{\bf 1},{\bf 1};{\bf 1},{\bf 2},{\bf 1})$ & & \\
& & & $2\times({\bf 2},{\bf 1},{\bf 1};{\bf 1},{\bf 1},{\bf 2})$ & & \\
& & & $2\times({\bf 1},{\bf 2},{\bf 1};{\bf 2},{\bf 1},{\bf 1})$ & & \\
& & & $2\times ({\bf 1},{\bf 1},{\bf 2};{\bf 2},{\bf 1},{\bf 1})$ & & \\
\hline
${\bf Z}_6$ & 2 & $[U(4) \otimes U(4)]_{99}\otimes$ 
& $({\bf 6},{\bf 1};{\bf 1},{\bf 1})$ & $14$ & $6$ \\
& &$[U(4) \otimes U(4)]_{55}$& $({\bf 1},{\bf 6};{\bf 1},{\bf 1})$ & & \\
& & & $({\bf 4},{\bf 4};{\bf 1},{\bf 1})$
& & \\
& & & $({\bf 1},{\bf 1};{\bf 6},{\bf 1})$
& & \\
& & & $({\bf 1},{\bf 1};{\bf 1},{\bf 6})$
& & \\
& & & $({\bf 1},{\bf 1};{\bf 4},{\bf 4})$
& & \\
& & & $2\times ({\bf 4},{\bf 1};{\bf 4},{\bf 1})$
& & \\
& & & $2\times ({\bf 1},{\bf 4};{\bf 1},{\bf 4})$
& & \\
\hline
& 4 & $[Sp(4) \otimes Sp(4)]_{99}$ & $({\bf 4},{\bf 4};
 {\bf 1},{\bf 1})$ & $14$ & $6$ \\
& &$[Sp(4) \otimes Sp(4)]_{55}$ & $({\bf 1},{\bf 1};{\bf 4},{\bf 4})$ & & \\
& &  & $2\times ({\bf 4},{\bf 1};{\bf 4},{\bf 1})$ & & \\
& & & $2\times ({\bf 1},{\bf 4};{\bf 1},{\bf 4})$ & & \\
\hline
\end{tabular}
\caption{The massless spectra of the six dimensional Type IIB orientifolds
on $T^4/{\bf Z}_N$ for $N=2,4,6$, and various values of $b$ 
(the rank of the $B$-field) worked out in [35].
The semi-colon in the column ``Charged Hypermultiplets'' separates $99$ and 
$55$ representations.}
\end{table}


\begin{references}

\bibitem{PS} G. Pradisi and A. Sagnotti, Phys. Lett. {\bf B216} (1989) 59;\\
M. Bianchi and A. Sagnotti, Phys. Lett. {\bf B247} (1990) 517; Nucl. Phys. 
{\bf B361} (1991) 539. 

\bibitem{GP} E.G. Gimon and J. Polchinski, Phys. Rev. {\bf D54} (1996) 1667.

\bibitem{GJ} E.G. Gimon and C.V. Johnson, Nucl. Phys. {\bf B477} (1996) 715;\\
A. Dabholkar and J. Park, Nucl. Phys. {\bf B477} (1996) 701.

\bibitem{bij} Z. Kakushadze, G. Shiu and S.-H.H. Tye, Phys. Rev. {\bf D58}
(1998) 086001.

\bibitem{BL} M. Berkooz and R.G. Leigh, Nucl. Phys. {\bf B483} (1997) 187.

\bibitem{Sagnotti} C. Angelantonj, M. Bianchi, G. Pradisi, A. Sagnotti and 
Ya.S. Stanev, Phys. Lett. {\bf B385} (1996) 96.

\bibitem{ZK} Z. Kakushadze, Nucl. Phys. {\bf B512} (1998) 221.

\bibitem{KS} Z. Kakushadze and G. Shiu, Phys. Rev. {\bf D56} (1997) 3686; 
Nucl. Phys. {\bf B520} (1998) 75.

\bibitem{Zw} G. Zwart, Nucl. Phys. {\bf B526} (1998) 378.

\bibitem{ibanez} G. Aldazabal, A. Font, L.E. Ib{\'a}{\~n}ez and G. Violero,
Nucl. Phys. {\bf B536} (1998) 29.

\bibitem{KST} Z. Kakushadze, G. Shiu and S.-H.H. Tye, 
Nucl. Phys. {\bf B533} (1998) 25.

\bibitem{blumen} R. Blumenhagen and A. Wisskirchen, Phys. Lett. {\bf B438}
(1998) 52.

\bibitem{223} Z. Kakushadze, Phys. Lett. {\bf B434} (1998) 269; 
Nucl. Phys. {\bf B535} (1998) 311; Phys. Rev. {\bf D58} (1998) 101901.

\bibitem{NP} Z. Kakushadze, hep-th/9904007.

\bibitem{pol} J. Polchinski, Phys. Rev. {\bf D55} (1997) 6423. 

\bibitem{blum} J.D. Blum, Nucl. Phys. {\bf B486} (1997) 34.

\bibitem{GJ1} E.G. Gimon and C.V. Johnson, Nucl. Phys. {\bf B479} (1996) 285.

\bibitem{berkooz} M. Berkooz, R.G. Leigh, J. Polchinski, J.H. Schwarz, 
N. Seiberg and E. Witten, Nucl. Phys. {\bf B475} (1996) 115.

\bibitem{paul} P.S. Aspinwall, Nucl. Phys. {\bf B496} (1997) 149.

\bibitem{SS} A. Sen and S. Sethi, Nucl. Phys. {\bf B499} (1997) 45.

\bibitem{MV} D.R. Morrison and C. Vafa, Nucl. Phys. {\bf B473} (1996) 74;
Nucl. Phys. {\bf B476} (1996) 437.

\bibitem{VW} C. Vafa and E. Witten, J. Geom. Phys. {\bf 15} (1995) 189.

\bibitem{anomalies}
M.B. Green, J.H. Schwarz and P.C. West, Nucl. Phys. {\bf B254} (1985) 327;\\
J. Erler, J. Math. Phys. {\bf 35} (1994) 1819.

\bibitem{aspin} P.S. Aspinwall, Phys. Lett. {\bf B357} (1995) 329.

\bibitem{Voisin} C. Voisin, in: Jorne{\'e}s de G{\'e}ometrie Alg{\'e}brique
d'Orsey, eds. A. Beauville {\em et al.}, Ast{\'e}risque, Vol. 218 (Soc. Math.
France, 1993) p.273.
 
\bibitem{Borcea} C. Borcea, in: Mirror Manifolds II, eds. B.R. Greene and S.-T.
Yau (International Press and AMS, 1997) p.717.

\bibitem{Asp} P.S. Aspinwall, Nucl. Phys. {\bf 460} (1996) 57.

\bibitem{Nik} V.V. Nikulin, in: Proc. ICM (Berkeley, 1986) p.654.

\bibitem{sen} A. Sen, Phys. Rev. {\bf D55} (1997) 7345; Nucl. Phys Proc. Suppl.
{\bf 68} (1998) 92.

\bibitem{SVW} S. Sethi, C. Vafa and E. Witten, Nucl. Phys. {\bf B480} (1996)
213. 

\bibitem{KLST} H. Kawai, D.C. Lewellen, J.A. Schwartz and S.-H.H. Tye,
Nucl. Phys. {\bf B299} (1988) 431.

\bibitem{cvetic} M. Cveti{\v c}, L. Everett, P. Langacker and J. Wang, 
hep-th/9903051.

\bibitem{new} Z. Kakushadze, to appear.

\bibitem{Bij} 
M. Bianchi, G. Pradisi and A. Sagnotti, Nucl. Phys. {\bf B376} (1992) 365.

\bibitem{unpublished} Z. Kakushadze and G. Shiu, unpublished.

\end{references}
\end{document}